\documentclass[preprint,prd,preprintnumbers,amssymb,superscriptaddress,nofootinbib]{revtex4}
\pdfoutput=1
\usepackage[english]{babel}
\makeatletter\AtBeginDocument{\let\@elt\relax}\makeatother
\usepackage{graphicx}
\usepackage{dcolumn}
\usepackage{bm}
\usepackage{slashed}
\usepackage{color}
\usepackage{amsmath}
\usepackage{multirow}
\usepackage{soul}
\usepackage{axodraw}
\usepackage{color}
\usepackage{enumitem}
\usepackage{hyperref}
\usepackage{array}
\usepackage{float}
\usepackage{lscape,graphicx} 
\usepackage{datenumber}
\usepackage{feynarts}
\usepackage{tikz}
\newcommand*\circled[1]{\tikz[baseline=(char.base)]{
            \node[shape=circle,draw,inner sep=1pt] (char) {#1};}}

\allowdisplaybreaks

\newcommand{\nn}{\nonumber}
\newcommand{\bea}{\begin{equation}\begin{array}{c}}
\newcommand{\eea}{\end{array}\end{equation}}
\newcommand{\ea}{\end{array}}

\newcommand{\beq}{\begin{equation}}
\newcommand{\eeq}{\end{equation}}
\newcommand{\bad}{\begin{array}{ccc}}

\begin{document}

\preprint{ULB-TH/15-21}
\preprint{TUM-HEP 1030/15}

\title{
 Probing the Inert Doublet Dark Matter Model \\ with Cherenkov Telescopes}

\author{Camilo  Garcia-Cely }
\affiliation{Service de Physique Th\'eorique, CP225, Universit\'e Libre de Bruxelles, Bld du Triomphe, 1050 Brussels, Belgium}

\author{Michael Gustafsson}
\affiliation{Institute for theoretical Physics - Faculty of Physics, Georg-August University G\"ottingen, Friedrich-Hund-Platz 1, D-37077 G\"ottingen, Germany}

\author{Alejandro Ibarra}
\affiliation{Physik-Department T30d, Technische Universit\"at M\"unchen, James-Franck-Stra\ss{}e, D-85748 Garching, Germany}

\begin{abstract}
We present a detailed study of the annihilation signals  of the inert dark matter doublet model in its high mass regime. Concretely, we study the prospects to observe gamma-ray signals of the model in current and projected Cherenkov telescopes taking into account the Sommerfeld effect and including the contribution to the spectrum from gamma-ray lines as well as from internal bremsstrahlung. We show that  present observations of the galactic center by the H.E.S.S. instrument are able to exclude regions of the parameter space that give the correct dark matter relic abundance. In particular, models with the charged and the neutral components of the inert doublet nearly degenerate in mass have strong gamma-ray signals. Furthermore, for dark matter particle masses above 1\,TeV, we find that the non-observation of the continuum of photons generated by the hadronization of the annihilation products typically gives stronger constraints on the model parameters than the sharp spectral features associated to annihilation into monochromatic photons and the internal bremsstrahlung process.  Lastly, we also analyze the interplay between  indirect and direct detection searches for this model, concluding that the prospects for the former are more promising. In particular, we find that  the upcoming Cherenkov Telescope Array will be able to probe a significant part of the high mass regime of the model.  
\end{abstract}

\maketitle
\tableofcontents

\section{Introduction}

Multiple observations strongly suggest that the Standard Model of particle physics should be extended by at least one additional particle, electrically neutral and colorless, and long-lived on cosmological time-scales, dubbed the dark matter (DM) particle~\cite{Bertone:2010zza,Bergstrom00,Jungman:1995df,Bertone:2004pz}. Among the many models that have been constructed over the last decades containing a DM particle, the Inert Doublet Model (IDM) stands out for its simplicity and for its rich phenomenology. 

The IDM~\cite{Deshpande:1977rw,Ma:2006km,Barbieri:2006dq} postulates the existence of a new scalar field $\eta$, with identical gauge quantum numbers as the Standard Model's Brout-Englert-Higgs scalar doublet (Higgs for short), and the invariance of the vacuum under a  $Z_2$ symmetry, under which $\eta$ is odd while all the Standard Model particles are even. These two simple assumptions have a number of implications. First, the $Z_2$ symmetry ensures that the doublet $\eta$ contains an absolutely stable particle which is a DM candidate. Second, the exotic doublet $\eta$ does not interact at tree level with any of the Standard Model fermions, hence the name ``inert''. The DM, nonetheless, interacts with the Standard Model via gauge interactions and via the quartic term in the scalar potential $|\eta|^2|\Phi|^2$, with $\Phi$ the Higgs doublet. These terms are of utmost importance in the phenomenology of the IDM, since they allow to generate a population of DM particles in the early Universe via thermal freeze-out and they induce potentially observable signals in direct and indirect DM searches \cite{Barbieri:2006dq,LopezHonorez:2006gr,Majumdar:2006nt,Gustafsson:2007pc,Agrawal:2008xz,Andreas:2009hj,Nezri:2009jd,Arina:2009um,Honorez:2010re,LopezHonorez:2010tb,Klasen:2013btp,Garcia-Cely:2013zga}, collider searches \cite{Barbieri:2006dq,Cao:2007rm,Lundstrom:2008ai,Dolle:2009ft,Miao:2010rg}, electroweak precision tests \cite{Barbieri:2006dq,Grimus:2008nb} and the Higgs diphoton decay rate \cite{Arhrib:2012ia,Swiezewska:2012eh,Krawczyk:2013jta, Goudelis:2013uca}.

Of particular interest in the IDM  is the scenario where the DM is entirely constituted by the lightest component of the doublet and where the observed DM abundance $\Omega h^2\simeq 0.12$~\cite{Ade:2015xua} is generated by thermal freeze-out of this particle. As is well known, this requirement implies for this model a DM particle mass either smaller than the W boson mass, $M_W\simeq 80$~GeV, or larger than $\sim 500$ GeV (see {\it e.g.} \cite{Barbieri:2006dq,LopezHonorez:2006gr,LopezHonorez:2010tb,Cirelli:2005uq,Hambye:2007vf,Hambye:2009pw}). In this paper we will concentrate in the latter mass window, and we will investigate the possibility of detecting gamma-ray signals generated by the annihilations of these DM particles in the Milky Way center. 

The DM induced gamma-ray flux is comprised of two main components. One component is the prompt gamma-rays, mainly generated by the hadronization of  massive gauge and Higgs bosons in the DM annihilation final states, and one lower energy component, consisting of photons of the interstellar radiation field that have been up-scattered to gamma-ray energies due to collisions with the energetic electrons and positrons produced in the annihilations. Since we are interested in the energy spectrum at the highest energies, we will neglect the latter contribution in what follows. Besides, the annihilation also produces sharp gamma-ray spectral features which, if observed, would strongly hint toward an exotic origin of this signal. So far, three different gamma-ray spectral features have been identified in DM scenarios: gamma-ray lines \cite{Srednicki:1985sf,Rudaz:1986db,Bergstrom:1988fp}, internal electromagnetic bremsstrahlung~\cite{Bergstrom:1989jr,Flores:1989ru,Beacom:2004pe,Bergstrom:2004cy,Bergstrom:2005ss,Bringmann:2007nk} and gamma-ray boxes~\cite{Ibarra:2012dw}. Notably, the three spectral features arise in the IDM: the gamma-ray lines arise from annihilations at the quantum loop level into $\gamma\gamma$ and $\gamma Z$, the internal bremsstrahlung signal arises from annihilation into $W^+W^-\gamma$ through the  t-channel exchange of the charged $Z_2$ odd scalars of the inert doublet, and gamma-ray boxes arise from the annihilation into a pair of Higgs bosons and their subsequent decay in flight into $\gamma\gamma$. Due to the small branching fraction of the process $h\rightarrow \gamma\gamma$, the gamma-ray box produced by the decay in flight of the Higgs boson is fainter than the other two spectral features and will not be considered here. 

Since the DM candidate in the IDM possesses a $SU(2)_L$ charge, weak gauge bosons could be exchanged between the non-relativistic particles in the initial state of the annihilation process. As argued in \cite{Hisano:2002fk, Hisano:2003ec,Hisano:2004ds}, for heavy DM particles the long-range interaction associated to the exchange of the weak gauge bosons can significantly distort the wave function of the initial state particles, therefore the correct description of the annihilation process must include non-perturbative effects, which generically lead to an enhancement of the annihilation cross section. This phenomenon, commonly known as Sommerfeld enhancement, can boost the annihilation signal by many orders of magnitude and has been proved to be pivotal in ruling out some well motivated DM scenarios, such as the Wino DM \cite {Chun:2012yt,Cohen:2013ama,Baumgart:2014vma,Ovanesyan:2014fwa,Baumgart:2014saa,Beneke:2014hja,Chun:2015mka} or the 5-plet minimal DM~\cite{Garcia-Cely:2015dda,Cirelli:2015bda} (assuming the DM density follows an Einasto profile in our Galaxy). The main goal of this paper is to investigate the prospects to observe signals in gamma-rays from the IDM in the high mass regime, including the Sommerfeld enhancement and inclu-ding not only the channels generating a continuum of gamma-rays, but also those generating sharp spectral features in the energy spectrum. 

The paper is organized as follows. In Section \ref{sec:DMHiggs} we present a brief overview of the Inert Doublet Model. In Section \ref{sec:DM-annihilation} we discuss the process of annihilation in the non-relativistic limit and we describe our non-perturbative approach to calculate the cross section. In Section \ref{sec:gammas} we calculate the expected gamma-ray flux and we confront the predictions of the model to limits on continuum gamma-ray fluxes and on  sharp gamma-ray spectral features.  In Sections \ref{sec:FvsC} and \ref{sec:DD}, we discuss the complementarity between direct detection experiments and gamma-ray instruments in probing the parameter space of the IDM and, lastly, in Section \ref{sec:conclusions} we present our conclusions. We also include three appendices discussing various theoretical and experimental constraints on the IDM, technical details of the Sommerfeld enhancement  and an estimation of its effect on relic density calculations in the early Universe.

\section{Dark Matter as an Inert Scalar}
\label{sec:DMHiggs}

The IDM is an extension of the Standard Model by one complex scalar field $\eta$, which is a singlet under $SU(3)_C$, doublet under $SU(2)_L$ and has hypercharge $1/2$. Furthermore, the model  postulates a discrete  $Z_2$ symmetry, preserved also in the electroweak vacuum, under which the Standard Model particles are even while the extra scalar doublet $\eta$ is odd. With this particle content, the Lagrangian can be cast as ${\cal L}={\cal L}_{\rm SM}+{\cal L}_\eta$, where ${\cal L}_{\rm SM}$ is the Standard Model Lagrangian including a potential for the Higgs doublet $\Phi$
\begin{equation}
{\cal L}_{\rm SM} \supset - m_1^2 \Phi^\dagger \Phi - \lambda_1 (\Phi^\dagger \Phi)^2 \;,
\end{equation}
and ${\cal L}_\eta$ is the most general $Z_2$ invariant Lagrangian involving the scalar doublet $\eta$ 
\begin{eqnarray}
{\cal L}_\eta&=&(D_\mu \eta)^\dagger  (D^\mu \eta) -m_2^2\eta^\dagger \eta  
- \lambda_2(\eta^\dagger \eta)^2
- \lambda_3(\Phi^\dagger \Phi)(\eta^\dagger \eta) \nonumber\\
&& - \lambda_4(\Phi^\dagger \eta)(\eta^\dagger \Phi) 
-\dfrac{1}{2} \left( \lambda_5(\Phi^\dagger \eta)(\Phi^\dagger \eta)+{\rm \text{\small h.c.}}\right),
\label{eq:L}
\end{eqnarray}
where $D_\mu$ denotes the covariant derivative. 

Due to the postulate that the $Z_2$ symmetry remains unbroken, only the Higgs doublet acquires an expectation value, therefore the doublets can be cast as
\begin{equation}
 \Phi  =\begin{pmatrix} G^+ \\  \frac{v_h+h+i G^0}{\sqrt{2}} \end{pmatrix}\;, \hspace{50pt}\eta= \begin{pmatrix} H^+ \\ \frac{1}{\sqrt2} \left( H^0 + i A^0 \right) \end{pmatrix}\;,
\label{fieldcompIDM}
\end{equation}
where $v_h\equiv \sqrt{-m_1^2/\lambda_1} \approx 246$\,GeV,  $G^0$ and $G^+$ provide the longitudinal components of the of the $Z$ and $W^+$ bosons through the Brout-Englert-Higgs mechanism and $h$ is the Standard Model Higgs.  On the other hand, the inert  sector consists of two charged states $H^\pm$, one CP-even neutral state $H^0$ and one CP-odd neutral state $A^0$. Furthermore, the preserved $Z_2$ symmetry ensures that the lightest particle in the inert doublet is absolutely stable and, if it is neutral, it constitutes a DM candidate; we will assume in what follows that this is the case for the  CP-even neutral state $H^0$.

The seven parameters in the scalar potential of the model can be recast in terms of the Higgs boson mass  $M_h \approx 125$\,GeV, the vacuum expectation value of the Higgs field $v_h$ and the DM mass $M_{H^0}$, together with the quartic couplings $\lambda_2$, $\lambda_3, \lambda_4$ and $\lambda_5$. The masses of the remaining inert scalars are given in terms of these parameters by
\begin{eqnarray}
M^2_{H^+} = M^2_{H^0} -\frac{1}{2}(\lambda_4+\lambda_5)v_h^2\,, \hspace{30pt}
M^2_{A^0} = M^2_{H^0} - \lambda_5 v_h^2 \;.
\label{masses}
\end{eqnarray}
These masses receive corrections at the quantum level. In particular, gauge interactions induce a splitting between the neutral and charged  scalar masses which is approximately $356$ MeV~\cite{Cirelli:2005uq}. However, this contribution can be compensated by an appropriate renormalization of the quartic couplings, resulting in a mass difference which can, in principle, be arbitrarily small. Therefore, in this paper we will take the mass differences among the inert scalars as free parameters, only constrained by, e.g., the perturbative condition $|\lambda_i|<4\pi$ in Eq.~\eqref{masses}. The parameters of the IDM are further constrained by the stability of the vacuum ~\cite{Gunion:2002zf,Gustafsson:2010zz} and by the unitarity of the S-matrix~\cite{Ginzburg:2004vp,Branco:2011iw}. We summarize the constrains we impose in  Appendix \ref{sec:AppendixA}, along with various experimental bounds coming from electroweak precision observables and collider searches.  

The inert scalar $H^0$ has the characteristics of a Weakly Interacting Massive Particle (WIMP) because its gauge interactions with the electroweak gauge bosons and its quartic coupling to the Higgs particle are of the required size to thermally produce $H^0$ of the right amount, via the freeze-out mechanism, to match the  observed DM content of our Universe $\Omega h^2\simeq 0.12$ \cite{Ade:2015xua}.

There are two allowed DM mass regimes for $H^0$ (see \textit{e.g.}, \cite{Barbieri:2006dq,LopezHonorez:2006gr,LopezHonorez:2010tb,Hambye:2007vf,Hambye:2009pw}). The first is for $H^0$ masses  below the $W$ boson mass, where the DM annihilates mostly into light fermions with a rate controlled by the size of the quartic couplings.\footnote{Three body annihilations of the type $H^0H^0 \to WW^* \to W f \bar{f}'$ are also important in some regions of the parameter space \cite{Honorez:2010re}.} For masses immediately above the $W$ boson threshold, the gauge couplings alone are large enough to suppress the $H^0$ abundance below the observed DM content (and given current experimental constraints, there is no longer room to avoid this conclusion by invoking destructive interference effects; see \cite{LopezHonorez:2010tb} and then, e.g., \cite{Gustafsson:2012aj}). Nonetheless, for $M_{H^0}\simeq 535$\,GeV, and vanishing quartic couplings, the annihilation rate into gauge bosons is sufficiently small to reproduce the observed DM density. For masses above $535$\,GeV, the correct relic density can  also be obtained if the
quartic couplings are appropriately chosen, because their effect is to increase the annihilation cross section. This forms the second, so called, high DM mass regime of the IDM.

The requirement of correct $H^0$ abundance thus implies larger and larger couplings as the DM mass increases. In fact, an upper limit on the DM mass can be derived by imposing perturbativity on the couplings. With the bounds of Appendix~\ref{sec:AppendixA}, and from the relic abundance calculation in Appendix~\ref{sec:AppendixC},  we find an upper limit of $M_{H^0} \lesssim 20 $ TeV.
In this paper we will investigate this high mass regime of the IDM  and in particular the possible signals in gamma-ray signals from annihilation of $H^0$ particles in the galactic center. To this end, we discuss the corresponding annihilation cross sections in the following section.

\section{Annihilation Cross Section into Gamma-rays}
\label{sec:DM-annihilation}

Various annihilation channels contribute to the gamma-ray flux in the IDM. The processes with the largest cross section are the tree-level two-body annihilations into $W^+ W^-$, $ZZ$ and $hh$, which generate a gamma-ray flux with a featureless energy spectrum. Processes arising at higher order in perturbation theory have a smaller cross section, however they can contribute significantly to the gamma-ray flux at energies close to the kinematical end-point of the annihilation and produce a sharp spectral feature in the energy spectrum. This is the case of the tree-level three-body annihilation into $W^+ W^- \gamma$, as well as the one-loop annihilations into $\gamma\gamma$ and $\gamma Z$. The former was studied in \cite{Garcia-Cely:2013zga} and leads to a bump close to the kinematical  end-point of the gamma-ray energy spectrum. This process is sizable  especially when $H^0$ and $H^\pm$ are relatively close in mass, which is typical in the high mass regime (see Eq.~\eqref{masses}).
The latter, on the other hand, were studied in \cite{Gustafsson:2007pc} in the low mass regime. The perturbative approach pursued in that paper, however, cannot be
applied to the high mass regime since for very large DM masses the predicted annihilation rates in the galactic center into $\gamma\,\gamma$ and $\gamma\,Z$ exceed the upper bound set by unitarity. In fact, for non-realtivistic velocities, the one-loop annihilation cross sections into photons are not suppressed by the DM mass but
rather by the W boson mass (see Ref.~\cite{GarciaCely:2014jha} for a detailed discussion). Similar shortcomings of the perturbative calculation have been pointed out for neutralino DM in the MSSM ~\cite{Bergstrom:1997fh,Bern:1997ng}, which were solved in Ref.~\cite{Hisano:2004ds} by pursuing a non-perturbative approach. 

To calculate the annihilation cross section into the various final states in the high mass regime of the IDM, we thus follow closely the 
formalism introduced in \cite{Hisano:2004ds,Hisano:2002fk, Cirelli:2007xd}. There they use the framework of non-relativistic field theory, which is well motivated by the fact that DM particles move slowly in our Galaxy.\footnote{A detailed description of the non-perturbative calculation of the annihilaton rate in the IDM can also be found in Ref.~\cite{GarciaCely:2014jha}.} The non-relativistic action is obtained by taking the non-relativistic limit description of the components of the inert doublet, which are assumed to be quasi-degenerate in mass,  and by integrating out the light particles, namely the Higgs boson and the gauge bosons. In this formalism, it is convenient to introduce auxiliary fields for the two-body states
\begin{equation}
s(\vec{x},\vec{r}) = \begin{pmatrix}s_{H^0H^0}(\vec{x},\vec{r})\\s_{A^0A^0}(\vec{x},\vec{r})\\s_{H^-H^+}(\vec{x},\vec{r})\end{pmatrix}
\end{equation}
where $s_{i}$  describes the wave functions of any of the pairs $i=(H^0,H^0)$, $(A^0,A^0)$ and $(H^-,H^+)$. Here $\vec{x}$ is the position of the center of mass of the system and $\vec{r}$ is the relative position vector for the pair of particles. In terms of these auxiliary fields, the two-body state effective action reads
\begin{eqnarray}
S_{\text{eff}} =
\int d^4x d^3r\,
 s^\dagger(\vec{x},\vec{r}) \left( i \partial_{x^0} + \dfrac{\nabla^2_x}{4 M_{H^0}}+ \dfrac{\nabla^2_r}{M_{H^0}}- V(r) + 2i \Gamma \delta(\vec{r}) \right)s(\vec{x},\vec{r}),
\label{ActionNRIDM}
\end{eqnarray}

\begin{figure}[t]
\centering
\hspace{-1.5cm}
\includegraphics[width=0.4\textwidth]{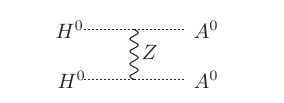}\hspace{-1.0cm}
\includegraphics[width=0.4\textwidth]{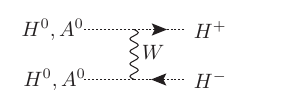}\hspace{-1.5cm}
\includegraphics[width=0.4\textwidth]{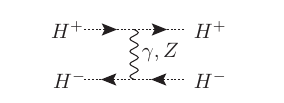}\hspace{-1.5cm}
\caption{\footnotesize Feynman diagrams contributing to $V_\text{Gauge}$(r)}
\label{fig:Vgauge}
\end{figure}

\begin{figure}[t]
\centering
\includegraphics[width=0.4\textwidth]{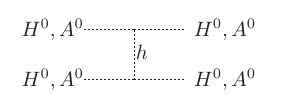}
\includegraphics[width=0.4\textwidth]{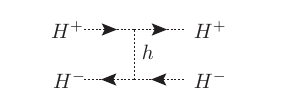}
\includegraphics[width=0.4\textwidth]{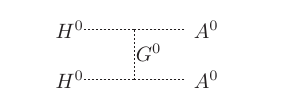}
\includegraphics[width=0.4\textwidth]{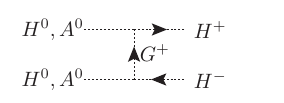}
\caption{\footnotesize Feynman diagrams contributing to $V_\text{Scalar}$(r)}
\label{fig:Vscalar}
\end{figure}

The matrix $V(r)$ represents a central potential consisting of three terms: one specifying the mass splittings among the pairs of particles and the other two describing Yukawa potentials induced by the non-relativistic  exchange of gauge bosons  (as shown in Fig.~\ref{fig:Vgauge}) and light scalars (as shown in Fig.~\ref{fig:Vscalar}). 
Thus
\begin{equation}
V(r) =  2\,\delta m  + V_\text{Gauge}(r) +V_\text{Scalar}(r), 
\label{potential}
\end{equation}
with
\begin{eqnarray}
\delta m &=& \begin{pmatrix}
0 & 0 & 0 \\
0 & M_{A^0}-M_{H^0} & 0 \\
0 & 0 & M_{H^+}-M_{H^0} \\
\end{pmatrix},\\
V_\text{Gauge}(r)&=&-\frac{g^2}{4\pi r}\begin{pmatrix}
0 & \frac{  e^{-M_Z r}}{4 c_W^2 }& \frac{  e^{-M_W r}}{2 \sqrt{2} }\\
\frac{ e^{-M_Z r}}{4 c_W^2 }   & 0 & \frac{e^{-M_W r}}{2 \sqrt{2} } \\
\frac{ e^{-M_W r}}{2 \sqrt{2} }  & \frac{ e^{-M_W r}}{2 \sqrt{2} } &  s_W^2 +\frac{(1-2 c_W^2)^2 e^{-M_Z r}}{4 c_W^2 }
\end{pmatrix},\\
V_\text{Scalar}(r)&=&-\frac{v_h^2}{8\pi r M_{H^0}^2}\begin{pmatrix}
\frac{(\lambda_3+\lambda_4+\lambda_5)^2}{2} e^{-M_h r} & \frac{\lambda_5^2 }{4} e^{-M_Z r}& \frac{(\lambda_4+\lambda_5)^2}{4\sqrt{2}} e^{-M_W r} \\
\frac{\lambda_5^2 }{4} e^{-M_Z r}  & \frac{(\lambda_3+\lambda_4-\lambda_5)^2}{2} e^{-M_h r} & \frac{(\lambda_4-\lambda_5)^2}{4 \sqrt{2}  }e^{-M_W r} \\
\frac{(\lambda_4+\lambda_5)^2}{4\sqrt{2}} e^{-M_W r} &   \frac{(\lambda_4-\lambda_5)^2}{4 \sqrt{2}  } e^{-M_W r} &  
\frac{\lambda_3^2}{2} e^{-M_h r}
\end{pmatrix}.
\end{eqnarray}
Here $s_W = \sin\theta_W$, $c_W= \cos\theta_W = M_W/M_Z$ and  $g = \sqrt{4\pi \alpha/s_W^2} \simeq 0.129$ with $\alpha \simeq 1/127$ being the fine structure constant.
In this formalism it is assumed that the splitting between the different pairs is negligible compared to the DM mass. 
 In fact, relatively small mass splittings $\delta M/M_{H^0}$ are unavoidable in this setup, because they are tied  to the electroweak breaking scale. Concretely, according to Eq.~\eqref{masses}, when $M_{H^0}$ is at the TeV scale, $\delta M  \sim \lambda v_h^2/M_{H^0}$ and  
therefore $\delta m_{ij} \ll M_{H^0}$.

Moreover, the action contains an absorptive -- or imaginary -- term, which takes into account that the two body states can annihilate into a Higgs pair or into two gauge bosons. This term is proportional to the matrix $\Gamma = \sum_f \Gamma^{(f)}$, where $f$ is any final state in which the pairs in the auxiliary field $s(\vec x,\vec r)$ can annihilate into. More concretely,
\begin{eqnarray}\hspace{-9pt}
\Gamma^{(f)}_{ij} = \frac{N_i N_j}{4M_{H_0}^2} \int
{\cal M} \left(i  \to f \right) {\cal M}^*\left(j \to f \right) (2\pi)^4 \delta^{(4)}\left(P_i-P_f \right)
\left(\prod_{a \in f } \frac{d^3 q_a}{(2\pi)^3 2 E_a} \right) \,,
\label{SEsigmav1}
\end{eqnarray}
\normalsize
where $P_i$ and $P_f$ are the total 4-momenta of the initial and final states, the $N_{i}$ are symmetry factors for the initial state particles with $N_{H^0H^0}= N_{A^0A^0}=1/\sqrt{2}$ and $N_{H^-H^+}=1$, and the integration is performed over the momentum $q_a$ of all  final state particles. For the 2-body final states
we find 
\begin{align}
\Gamma^{(\gamma\gamma)}
&= \frac{1}{2} \tan^2(2\theta_W) \Gamma_{Z\gamma}
= \frac{e^4}{128\pi{M_{H^0}}^2}
\left(
\begin{array}{ccc}
 0 & 0 & 0 \\
 0 & 0 & 0 \\
 0 & 0 & 16 \\
\end{array}
\right),\label{Gammagamma}\\
\Gamma^{(ZZ)} &= \frac{g^4}{128 c_W^4 \pi{M_{H^0}}^2}
\left(
\begin{array}{ccc}
 \frac{1}{2 } & \frac{1}{2 } & \frac{\left(1-2 c_W^2\right)^2 }{\sqrt{2} } \\
 \frac{1}{2 } & \frac{1}{2 } & \frac{\left(1-2 c_W^2\right)^2 }{\sqrt{2} } \\
 \frac{\left(1-2 c_W^2\right)^2 }{\sqrt{2} } & \frac{\left(1-2 c_W^2\right)^2 }{\sqrt{2} }
   & \left(1-2 c_W^2\right)^4  \\
\end{array}
\right)\nn\\
&+
\Gamma^{(SS)} \left(\lambda_3+\lambda_4-\lambda_5,\lambda_3+\lambda_4+\lambda_5,\lambda_3\right),\label{GammaZZ}\\
\Gamma^{(W^+W^-)} &= \frac{g^4}{128\pi{M_{H^0}}^2}
\left(
\begin{array}{ccc}
 1 & 1 & \sqrt{2}  \\
 1 & 1 & \sqrt{2}  \\
 \sqrt{2}  & \sqrt{2} & 2  \\
\end{array}
\right)+2\,\Gamma^{(SS)} \left(\lambda_3,\lambda_3,\lambda_3+\lambda_4\right), \label{GammaWW}\\
\Gamma^{(hh)} &=\Gamma^{(SS)} \left(\lambda_3+\lambda_4+\lambda_5,\lambda_3+\lambda_4-\lambda_5,\lambda_3\right)\label{Gammahh}\,,
\end{align}
where we introduce for convenience the following matrix 
\begin{eqnarray}
\Gamma^{(SS)} \left(l_1,l_2,l_3\right) &\equiv& \frac{1}{128\pi{M_{H^0}}^2}
\left(
\begin{array}{ccc}
  l_1^2 &  l_1 l_2 &  \sqrt{2} l_1 l_3 \\
  l_1 l_2 &  l_2^2 &  \sqrt{2} l_2 l_3 \\
  \sqrt{2} l_1 l_3 &  \sqrt{2} l_2 l_3 & 2 l_3^2 \\
\end{array}
\right)\,.
\end{eqnarray}
For the internal bremsstrahlung process, corresponding to the final states $WW\gamma$, we directly use Eq.~\eqref{SEsigmav1}, before integrating on the photon energy (see the Appendix of Ref.~\cite{Garcia-Cely:2013zga} for details).

These matrices are of interest here because they allow to calculate the annihilation $s$-wave cross section in the final state $f$ by means of the formula
\begin{equation}
\sigma v \left(i   \to f\right) \Big|_\text{s-wave} = \frac{1}{N_{i }^2} (d\,\Gamma^{(f)}d^\dagger)_{ii},
\label{SEsigmavText}
\end{equation}
where the matrix $d$ are the Sommerfeld enhancement factors that can be calculated by solving the Schr\"odinger equation associated to the potential~\eqref{potential}, as described in detail in Appendix \ref{sec:AppendixB}.

\begin{figure}[t]
\begin{center}
\includegraphics[width=0.49\textwidth]{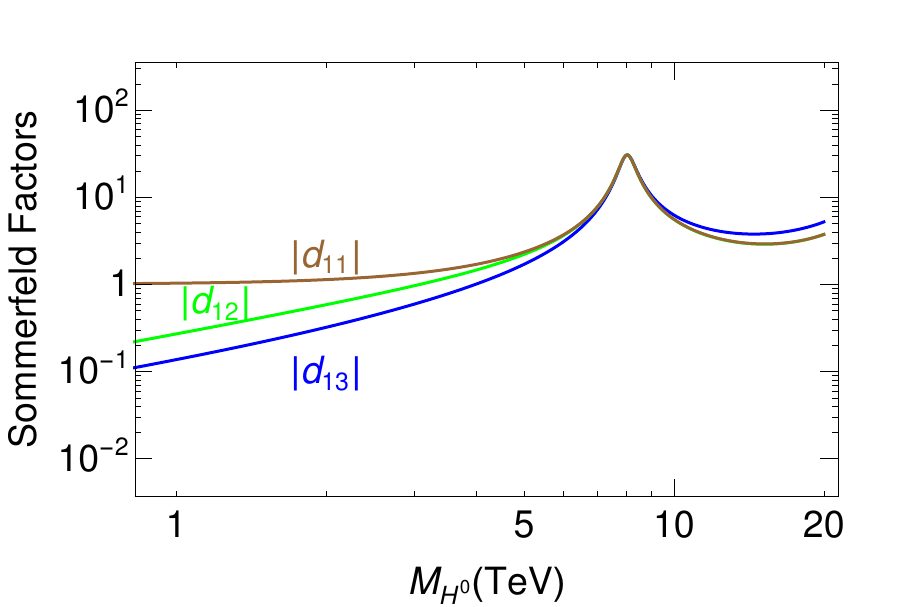}
\includegraphics[width=0.49\textwidth]{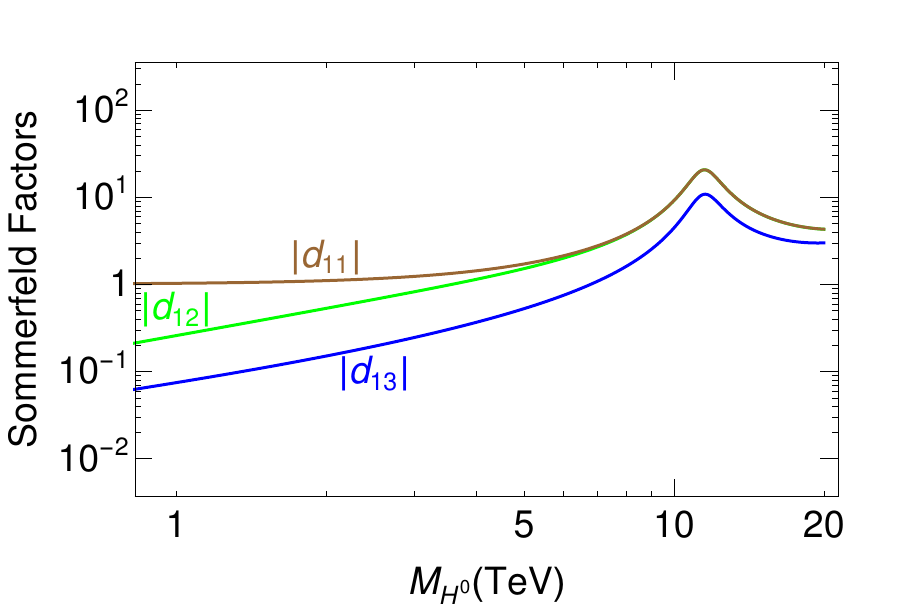}
\caption{Example of the Sommerfeld enhancement factors  for annihilating DM particles with a relative velocity of $v=2\times 10^{-3}$,  $\lambda_3 = \lambda_5 = 0$ and $M_{H^+}-M_{H^0} = 1$\,GeV (left panel) or $M_{H^+}-M_{H^0} = 10$\,GeV (right panel).}
\label{fig:GaugeRes}
\end{center}
\end{figure}

For DM annihilation, Eq.~(\ref{SEsigmavText}) can be cast as
\small
\begin{eqnarray}\hspace{-9pt}
\sigma v \left(H^0 H^0 \to f\right) \Big|_\text{s-wave} = \frac{1}{4M^2_{H^0}} \int \left(\prod_{a \in f } \frac{d^3 q_a}{(2\pi)^3 2 E_a} \right) (2\pi)^4 \delta^4\left(p_{H^0} +p'_{H^0}-{\scriptstyle\sum_{a \in f}} q_a\right) \nonumber\\
\times\Bigg| {d_{11}\;\cal M} \left(H^0H^0 \to f\right)+d_{12}\;{\cal M} \left(A^0A^0 \to f\right)+\sqrt{2} d_{13}\;{\cal M} \left(H^+H^- \to f\right)\Bigg|^2 \,.
\label{SEsigmav2}
\end{eqnarray}
\normalsize
The quantities $d_{11}$, $d_{12}$ and $d_{13}$ are therefore  interpreted as non-perturbative enhancement factors that account for the long range interactions between the annihilating DM particles due to the exchange of gauge and  Higgs bosons in the non-relativistic limit. As an example, we show in Fig.~\ref{fig:GaugeRes} the absolute value of $d_{11}$, $d_{12}$ and $d_{13}$ as a function of the DM mass, for the case $\lambda_3=\lambda_5=0$ and $\lambda_4$ chosen so that the mass splitting between the charged and the neutral component is 1\,GeV (left panel) and 10\,GeV (right panel)\footnote{An error in the previous versions of Fig.~\ref{fig:GaugeRes} was corrected. This did not affect any other result of the paper. }. We find that, for masses below approximately 2~TeV, the inclusion of these factors in the calculation is irrelevant; however, once the DM mass increases, the enhancement factors dramatically affect the annihilation cross sections in Eq.~(\ref{SEsigmav2}). Furthermore, we find a resonance, which moves to higher DM masses as the mass splitting between the charged and the neutral component is increased. This is in agreement with what was found in Ref.~\cite{Hisano:2004ds} for neutralino DM. 

In order to study the impact of the Sommerfeld enhancement in the IDM, we performed a scan over the five dimensional parameter space of the DM sector following the procedure described below. Considering,  as observed in Fig.~\ref{fig:GaugeRes}, that scalar mass splittings are essential quantities, instead of taking $\lambda_4$ and $\lambda_5$ as parameters of the scan, we let the relative mass splittings $(M_{H^+}-M_{H^0})/M_{H^0}$ and $(M_{A^0}-M_{H^0})/M_{H^0}$ vary logarithmically in between $10^{-5}$ and $1$. Then using Eq.~\eqref{masses}, we solve for $\lambda_4$ and $\lambda_5$ and we discard  points whose magnitude is greater than $4\pi$. In contrast, the quartic couplings $\lambda_2$ and $\lambda_3$, which do not lead to any mass splitting, are randomly sampled  on a linear scale. Finally, we take $M_{H^0}$ in-between 0.5 and 20\,TeV  and impose the remaining constraints of Appendix~\ref{sec:AppendixA}.
We then calculate the relic abundance of $H^0$ for each model by means of micrOMEGAs 3.1~\cite{Belanger:2013oya} and  require that it is in agreement with the observed value $\Omega_{\text{DM}} h^2=0.1199\pm 0.0027$ within a 30\% range. As discussed in Appendix \ref{sec:AppendixC}, 30\% is the error in the relic density calculation that  can be expected from our approximation of not accounting for the Sommerfeld effect in the early Universe.

\begin{figure}[t]
\begin{center}
\includegraphics[width=0.49\textwidth]{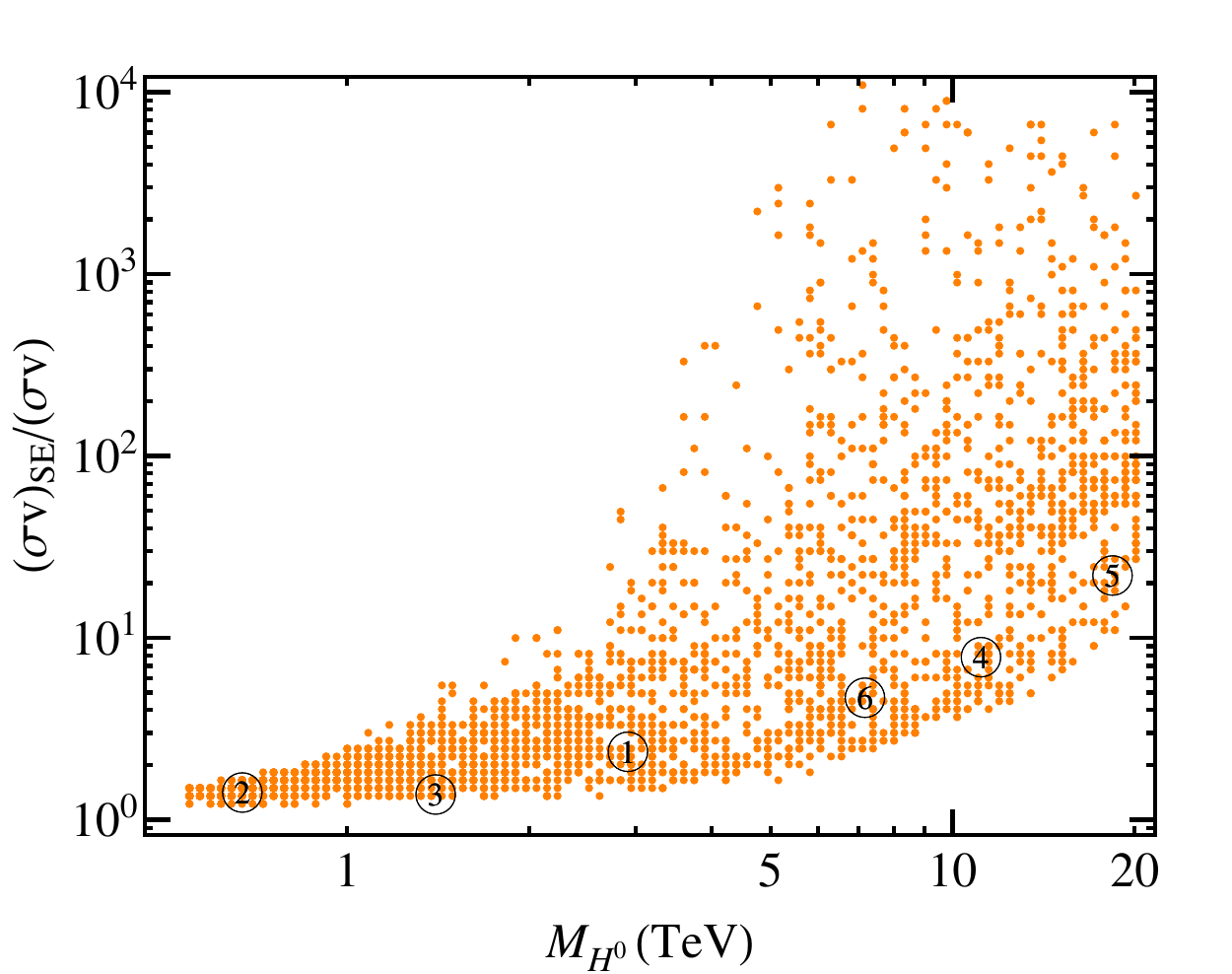}\\
\includegraphics[width=0.49\textwidth]{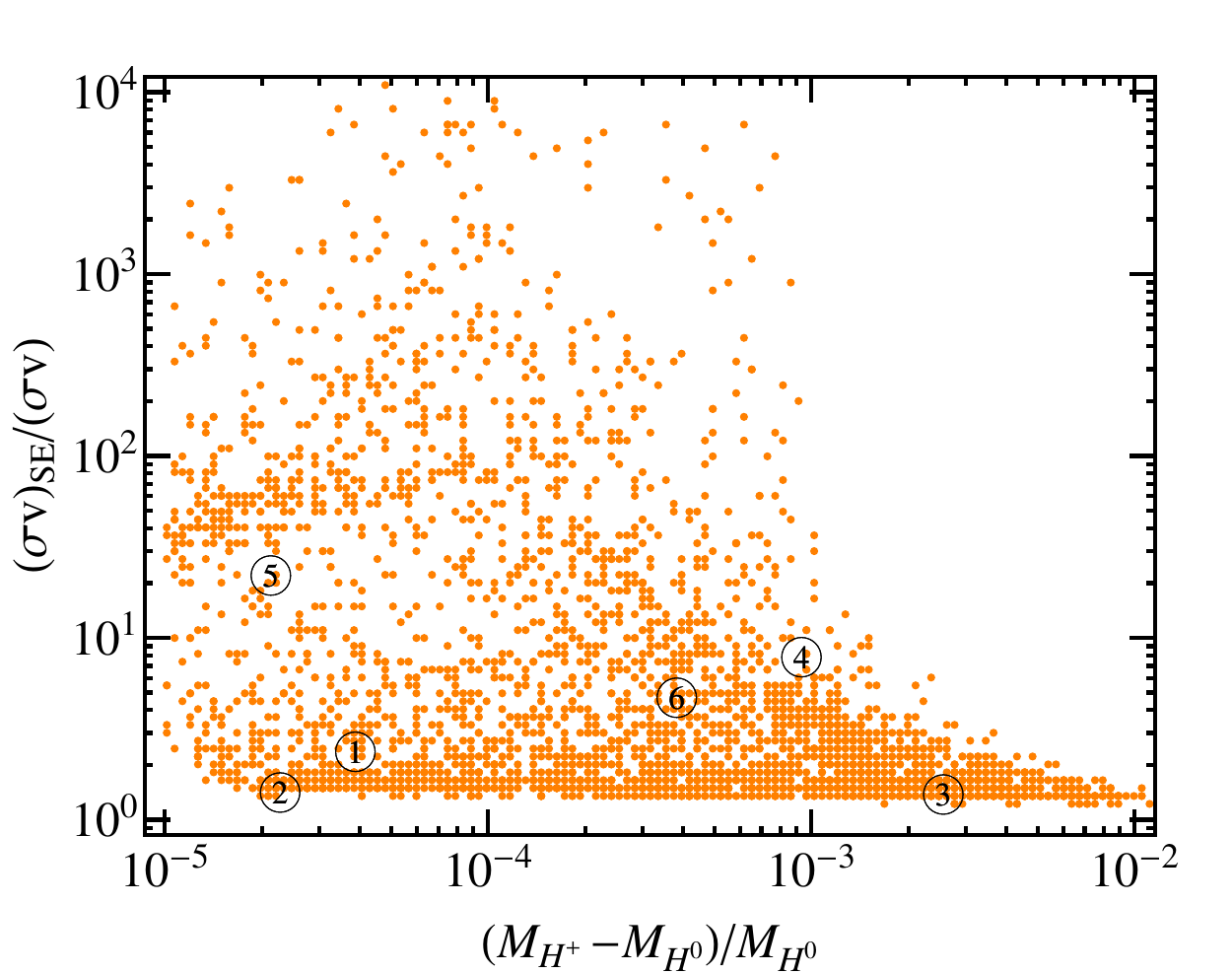}
\includegraphics[width=0.49\textwidth]{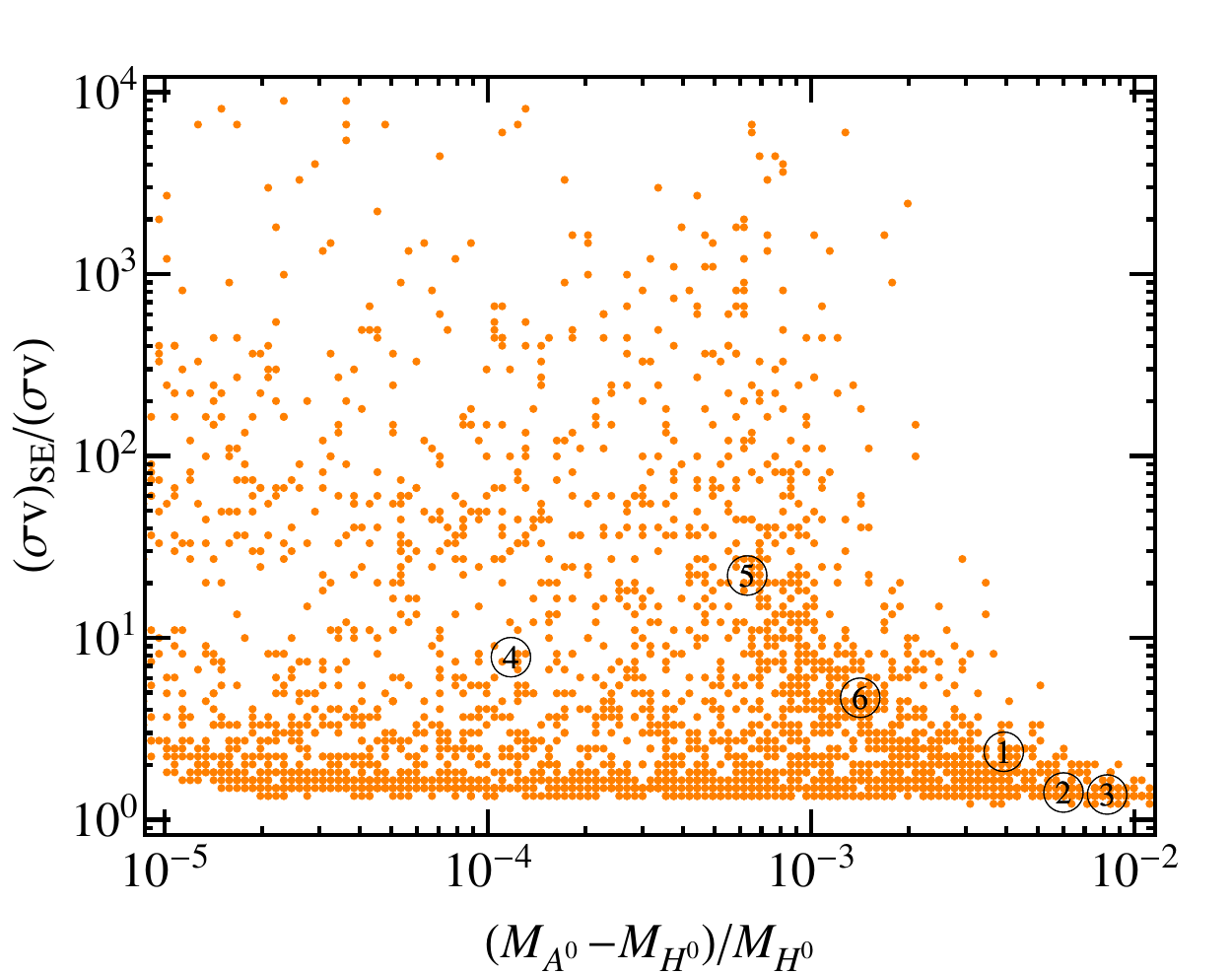}
\caption{\small Impact of the Sommerfeld enhancement on the total annihilation cross section for a random sample of points of the viable parameter space, as described in the text. We also show the benchmark points of Table \ref{table:Spectra}. Here the DM velocity is  $v=2\times 10^{-3}$.}
\label{fig:SEboost}
\end{center}
\end{figure}

Assuming a relative velocity between annihilating DM particles of $v=2\times 10^{-3}$, for each model point of the scan we calculate the enhancement factors $d_{11}$, $d_{12}$ and $d_{13}$. Then, we calculate the total annihilation cross section, i.e.\ the sum of the  $H^0 H^0\rightarrow W^+W^-,~ZZ,~hh$ cross sections. The resulting enhancement  on the total annihilation cross section  is shown in Fig.~\ref{fig:SEboost}
as a function of the DM mass (upper panel) and the relative mass splittings of the inert scalars (lower panel). In the plot we observe two facts. On the one hand, the Sommerfeld effect is only relevant for masses in the TeV scale, as expected, and its importance increases for masses close to 20\,TeV. On the other hand, the smaller the mass splittings the greater the Sommerfeld effect. 

The enhancement of the cross section in the channels producing gamma-rays in the final state has important implications for the indirect searches of the inert doublet DM at gamma-ray telescopes, as we discuss in the next section.

\section{Gamma-ray Signals of the IDM}
\label{sec:gammas}

The DM induced gamma-ray signal from a given sky region is
\begin{eqnarray}
\frac{d\phi_\gamma}{dE_\gamma } = \frac{\bar{J}} {8 \pi M^2_{H^0}}
\frac{d(\sigma  v)_\gamma}{dE_\gamma}\,.
\label{fluxeqIDM}
\end{eqnarray}
The astrophysical $\bar{J}$-factor is here the line-of-sight integral over the squared DM density $\rho_{H^0}$, averaged over the observed solid angle $\Delta\Omega$, 
\begin{eqnarray}
\bar{J} = \frac{1}{\Delta \Omega} \int_{\Delta \Omega } \int_{\text{l.o.s.}} \rho^2_{H^0} \,ds \, d\Omega \,.
\end{eqnarray}
The $d(\sigma v)_\gamma/dE_\gamma$ is comprised of three different components: first, the fairly featureless spectrum generated in the decay and fragmentation of the gauge and Higgs bosons, second, the gamma-ray lines produced by the monochromatic photons emitted in the processes $H^0H^0\to\gamma\gamma$ and $H^0H^0\to\gamma Z$, and third, the virtual internal bremsstrahlung signal from $H^0 H^0\rightarrow W^+ W^- \gamma$. Concretely, the  inclusive differential cross section into photon is
\begin{eqnarray}
\frac{d (\sigma v)_\gamma}{dE_\gamma}  =  \sum_{f \in \text{two-body}} \!\!\!\!\! \sigma v(H^0 H^0\to f) \frac{dN^f_\gamma}{dE_\gamma} \;\;\; + \;  \frac{d\,\sigma v (H^0 H^0 \to WW \gamma)}{dE_\gamma},
\label{eq:dNdE}
\end{eqnarray}
where $dN^f_\gamma/dE_\gamma$ is the photon multiplicity associated to the two body final states $f$. When $f$ is a electroweak or Higgs boson pair, we use the parametrization $dN^\gamma/dE_\gamma = dN^\text{frag}_\gamma/dE_\gamma = \frac{0.73}{M_{H^0}}\,x^{1.5}\,e^{-7.8x}$ with $x=E_\gamma/M_{H^0}$ \cite{Bergstrom:1997fj}. For the $\gamma\gamma$ and $\gamma Z$ final states, $dN^f_\gamma/dE_\gamma$ is a delta function at $E_\gamma = M_{H^0}$ and $M_{H^0}-\frac{M_Z^2}{4M_{H^0}}$, respectively.  The $dN^\text{frag}_\gamma/dE_\gamma$ does not account for initial internal bremsstrahlung contributions, $WW\gamma$, which is instead explicitly included by the last term of Eq.~\eqref{eq:dNdE}. 

The relative strength of each of these components in the total gamma-ray flux strongly depends on the concrete choice of the parameters of the model. In order to assess the prospects to observe annihilation signals of the IDM, we have calculated the predicted gamma-ray flux for all the viable models of our parameter scan from section~\ref{sec:DM-annihilation}. We include  the gamma-ray contributions from the final states $W^+W^-$, $ZZ$, $hh$, $\gamma\gamma$, $\gamma Z$ and $W^+W^-\gamma$ and take into account the Sommefeld enhancement for each of these channels\footnote{All other annihilation channels are always subdominant, with the top-quark channel potentially reaching a ratio of up to 5\% for our lowest DM masses.}, as described in Section \ref{sec:DM-annihilation}. In order to better illustrate results, we also selected six benchmark model points (BMPs) displaying qualitatively different spectra. The parameters corresponding to each of these points as well as their predicted gamma-ray energy spectra are shown in Table \ref{table:Spectra}. 
\begin{table}[ht!]
\centering
\renewcommand{\arraystretch}{0.79}
\begin{tabular}{|c|c|c|} \hline
{\bf Spectrum} & \multicolumn{2}{|c|}{\bf Benchmarks} 
\\\hline
%
& \scriptsize {\bf BMP 1:}  $M_{H^0}=2.91$ TeV
& \scriptsize {\bf BMP 2:}  $M_{H^0}=0.67$ TeV
\\
&\scriptsize $\lambda_3=0.120\;\;\lambda_4=1.078\;\;\lambda_5=-1.100$
&\scriptsize$\lambda_3=0.090\;\;\lambda_4=0.089\;\;\lambda_5=-0.090$
 \\
&
\multirow{4}{*}{
\includegraphics[trim=0.05cm 0cm 1.3cm 1.1cm,clip,width=6.1cm]{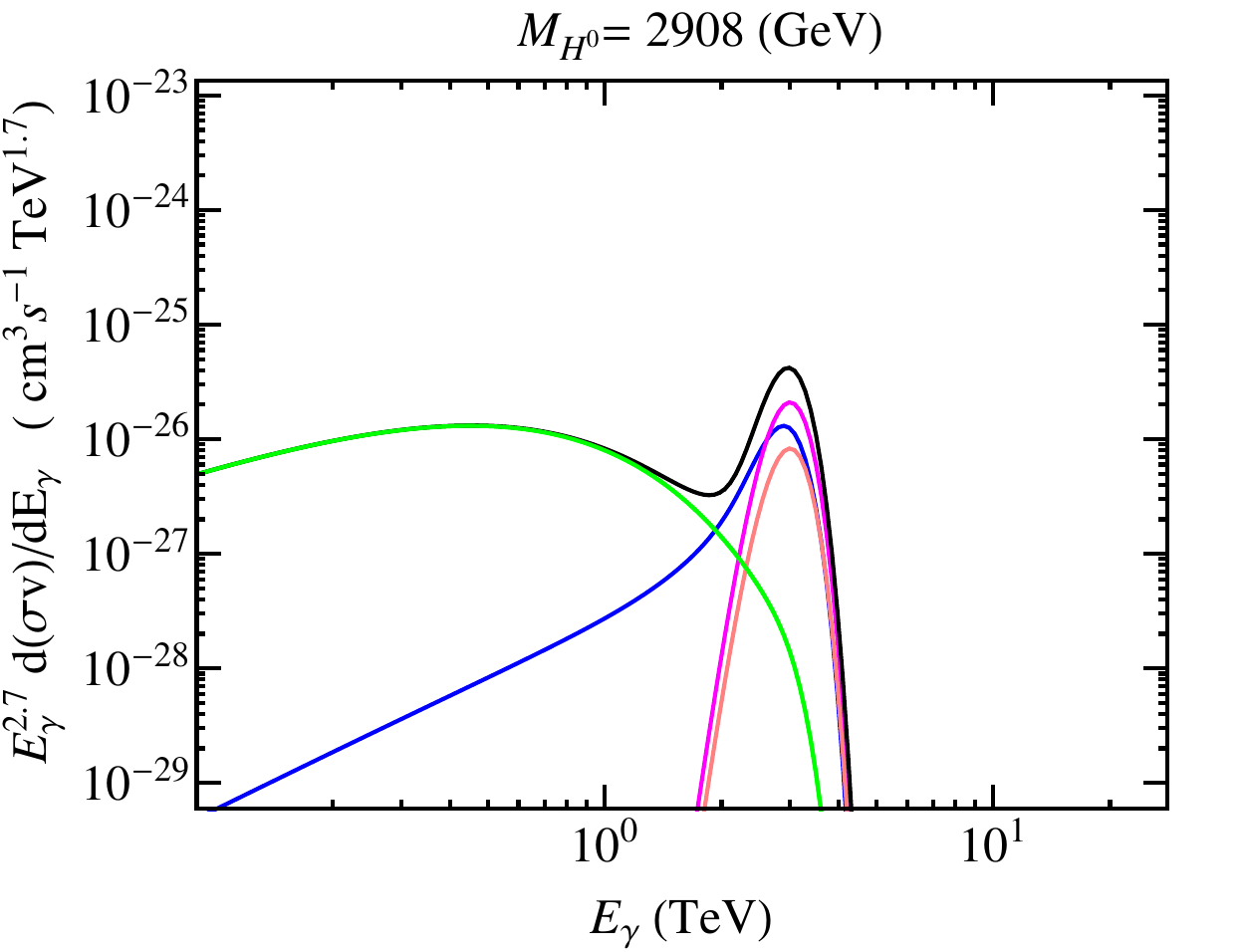}
}
&
\multirow{4}{*}{
\includegraphics[trim=0.05cm 0cm 1.3cm 1.1cm,clip,width=6.1cm]{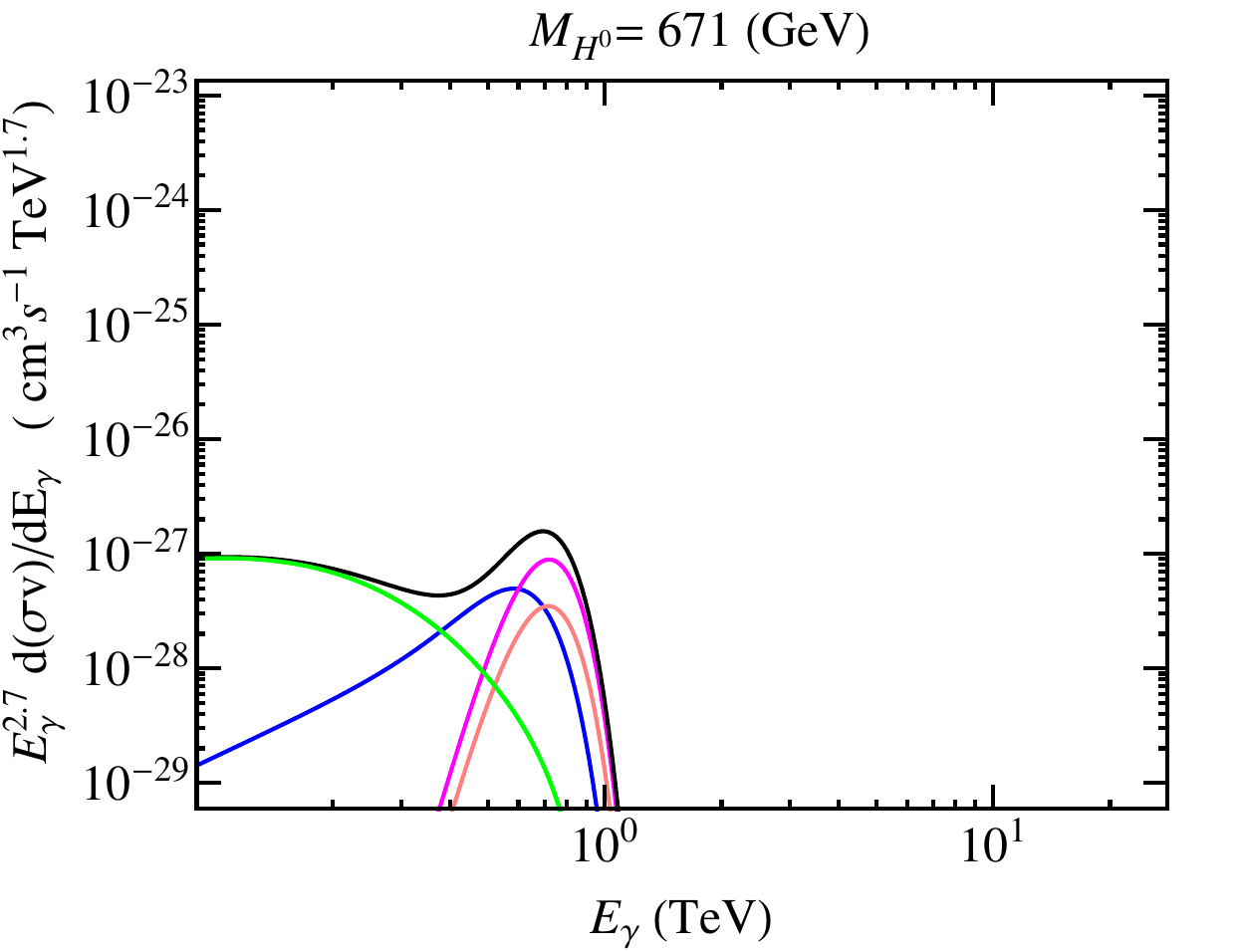}
}\\
&&\\
&&\\
Dominated&&\\
by $\gamma$ lines&&\\
&&\\
&&\\
&&\\\hline
&\scriptsize {\bf BMP 3:} $M_{H^0}=1.40$ TeV
& \scriptsize {\bf BMP 4:} $M_{H^0}=11.14$ TeV
\\
&\scriptsize$\lambda_3=-0.370\;\;\lambda_4=0.202\;\;\lambda_5=-0.533$
&\scriptsize$\lambda_3=6.870\;\;\lambda_4=-7.146\;\;\lambda_5=-0.481$
 \\
&
\multirow{4}{*}{
\includegraphics[trim=0.0cm 0cm 1.3cm 1.1cm,clip,width=6.2cm]{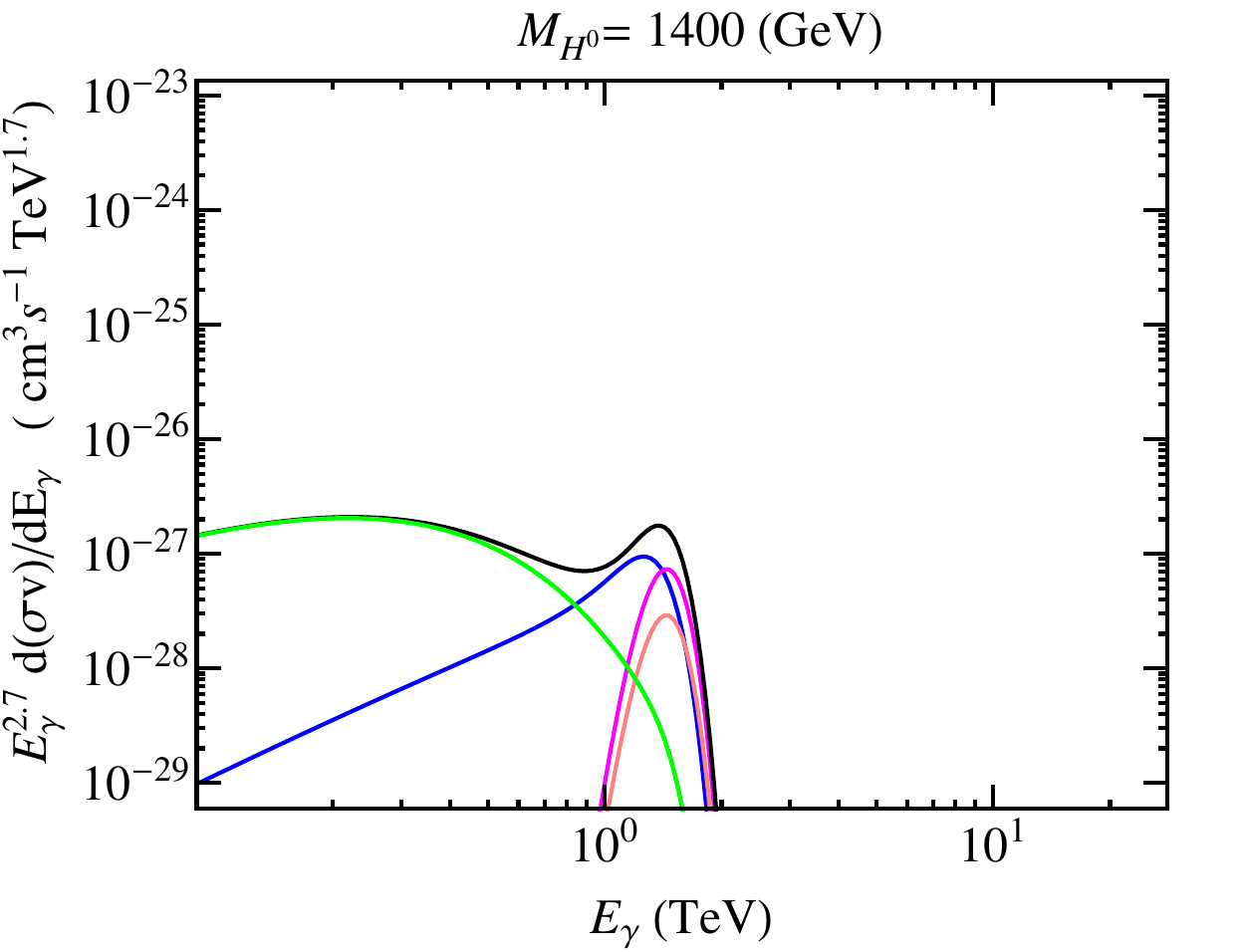}
}
&
\multirow{4}{*}{
\includegraphics[trim=0.0cm 0cm 1.3cm 1.1cm,clip,width=6.2cm]{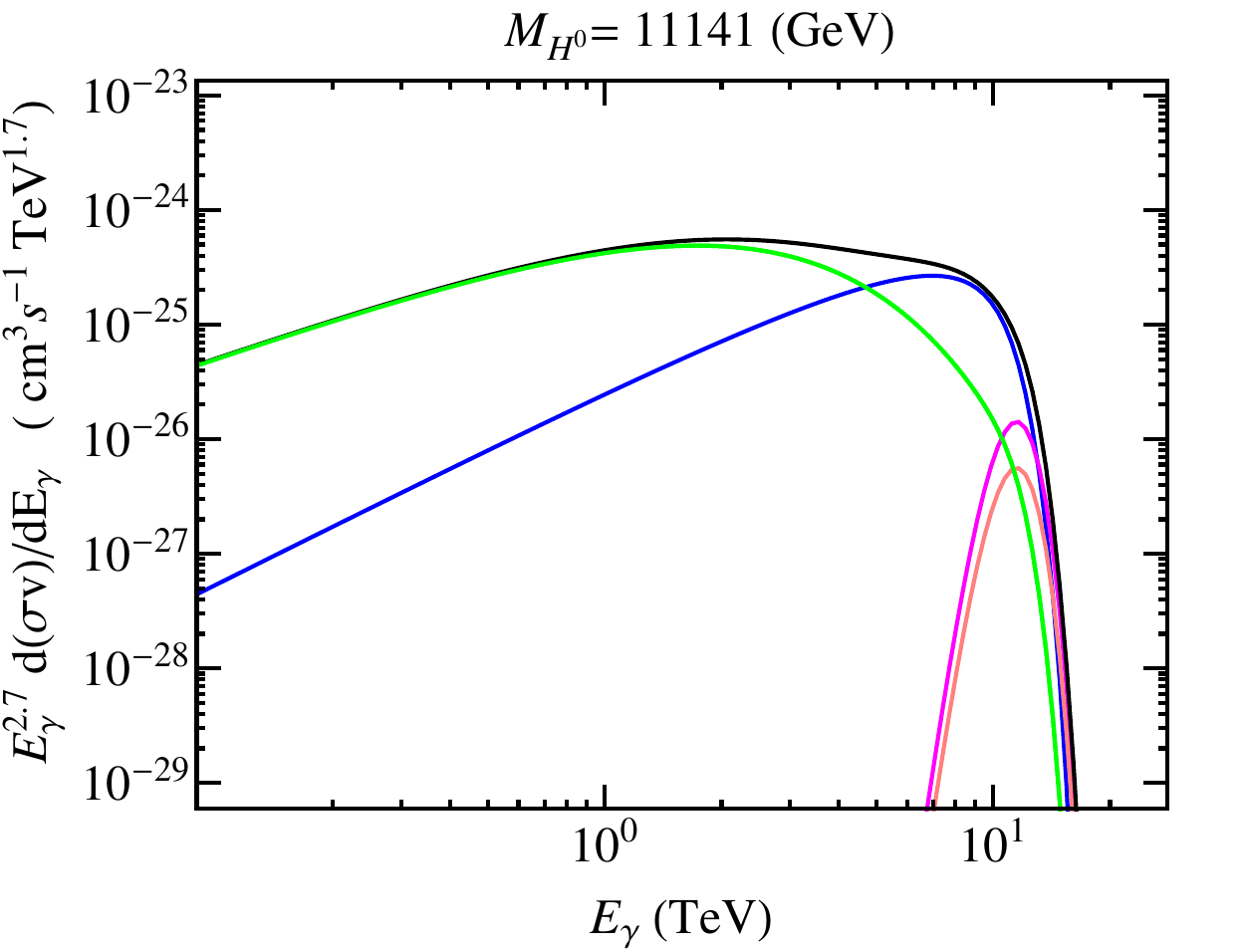}
}\\
&&\\
&&\\
Dominated&&\\
by VIB&&\\
&&\\
&&\\
&&\\\hline
& \scriptsize {\bf BMP 5:} $M_{H^0}=18.38$ TeV 
& \scriptsize {\bf BMP 6:} $M_{H^0}=7.17$ TeV 
\\
&\scriptsize$\lambda_3=0.900\;\;\lambda_4=6.572\;\;\lambda_5=-7.046$
&\scriptsize$\lambda_3=3.130\;\;\lambda_4=1.101\;\;\lambda_5=-2.399$
\\
&
\multirow{4}{*}{
\includegraphics[trim=0.0cm 0cm 1.3cm 1.1cm,clip,width=5.8cm]{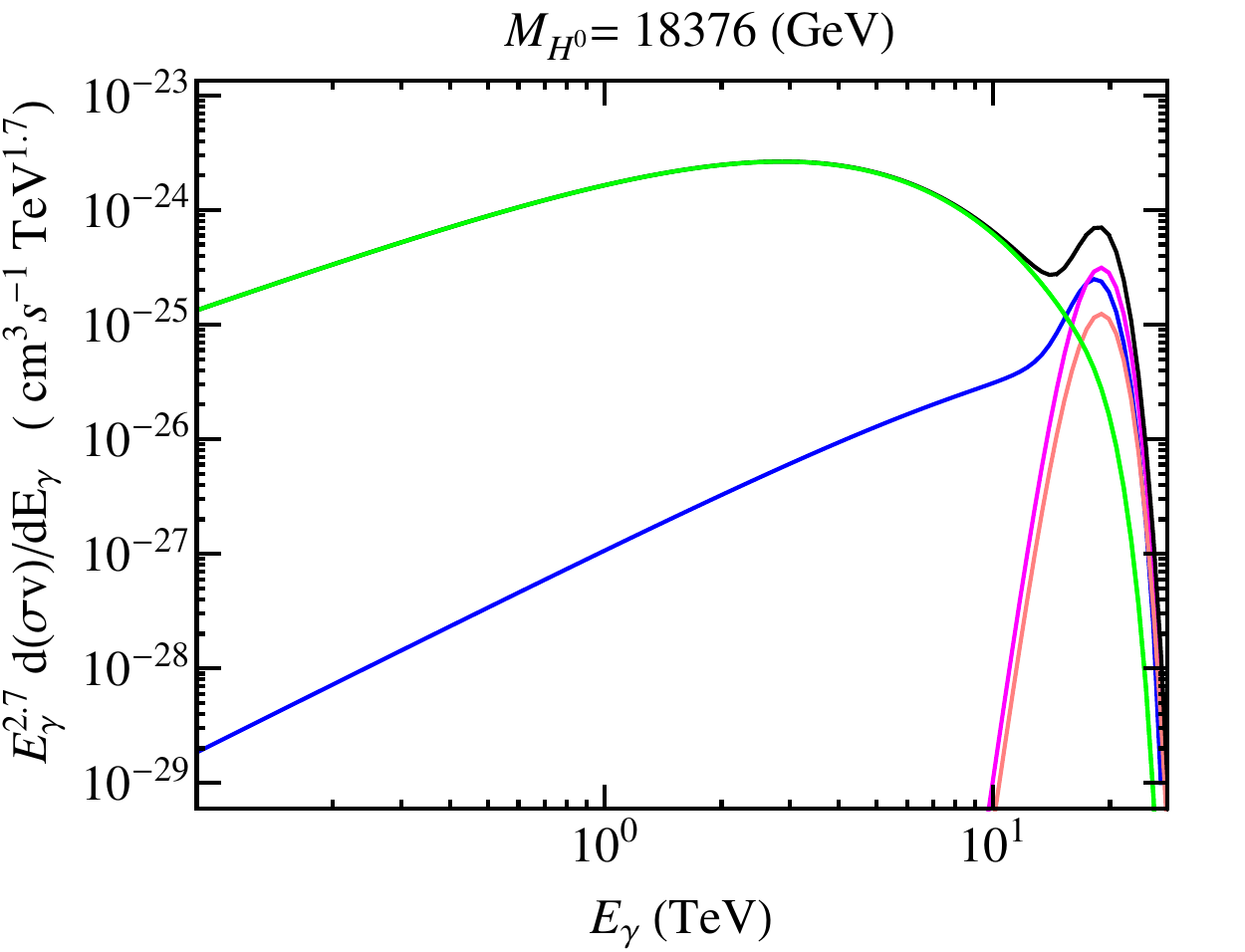}
}
&
\multirow{4}{*}{
\includegraphics[trim=0.0cm 0cm 1.3cm 1.1cm,clip,width=5.8cm]{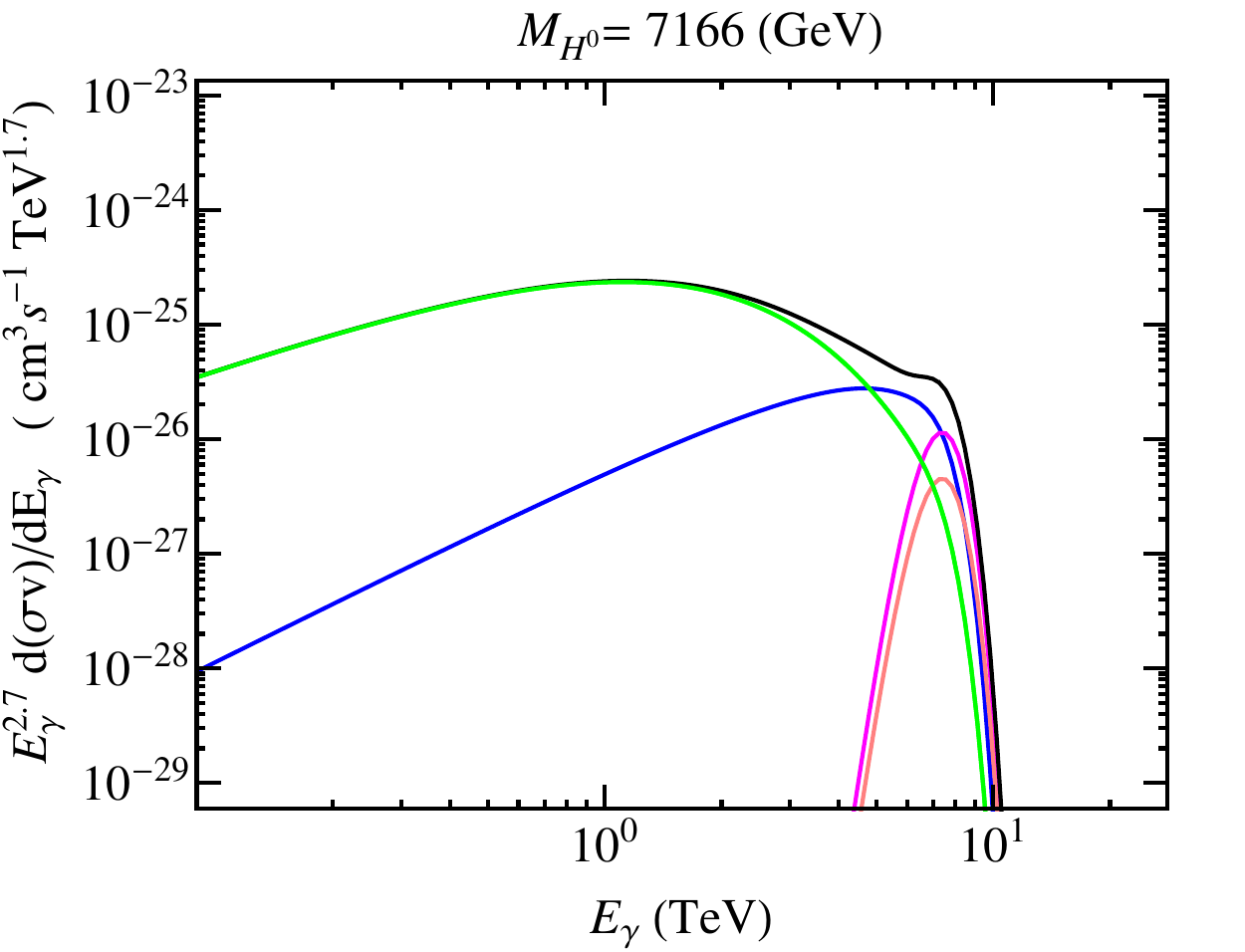}
}\\
&&\\
&&\\
Dominated&&\\
by the &&\\
continuum&&\\
&&\\
&&\\\hline
\end{tabular}
\caption{\small Contributions to the differential cross section from continuum photons (green line), $\gamma\gamma$ (magenta line) and $\gamma Z$ (pink line) and virtual internal bremsstrahlung [VIB] (blue line), as well as the total differential cross section (black line) for our selected six benchmark points in the viable parameter space. The spectra have been convoluted with a Gaussian detector response characterized by a 10\% energy resolution.}
\label{table:Spectra}
\end{table}
Benchmark points BMP1 and BMP2 produce a very intense gamma-ray line, BMP3 and BMP4 produce a significant virtual bremsstrahlung signal, while BMP5 and BMP6 produce an intense continuum. In the plots, the contributions of the virtual internal bremsstrahlung (VIB), the continuum part, and the $\gamma\gamma$ and $\gamma Z$ monochromatic lines are shown, respectively, in blue, green, magenta and pink.  Considering that the total gamma-ray flux observed by H.E.S.S. telescope falls roughly as $E^{-2.7}$ \cite{Abramowski:2013ax}, our spectra have been multiplied by $E^{2.7}$ in order to better appreciate their features at the highest energies.

\begin{figure}[t]
\begin{center}
\includegraphics[width=0.49\textwidth]{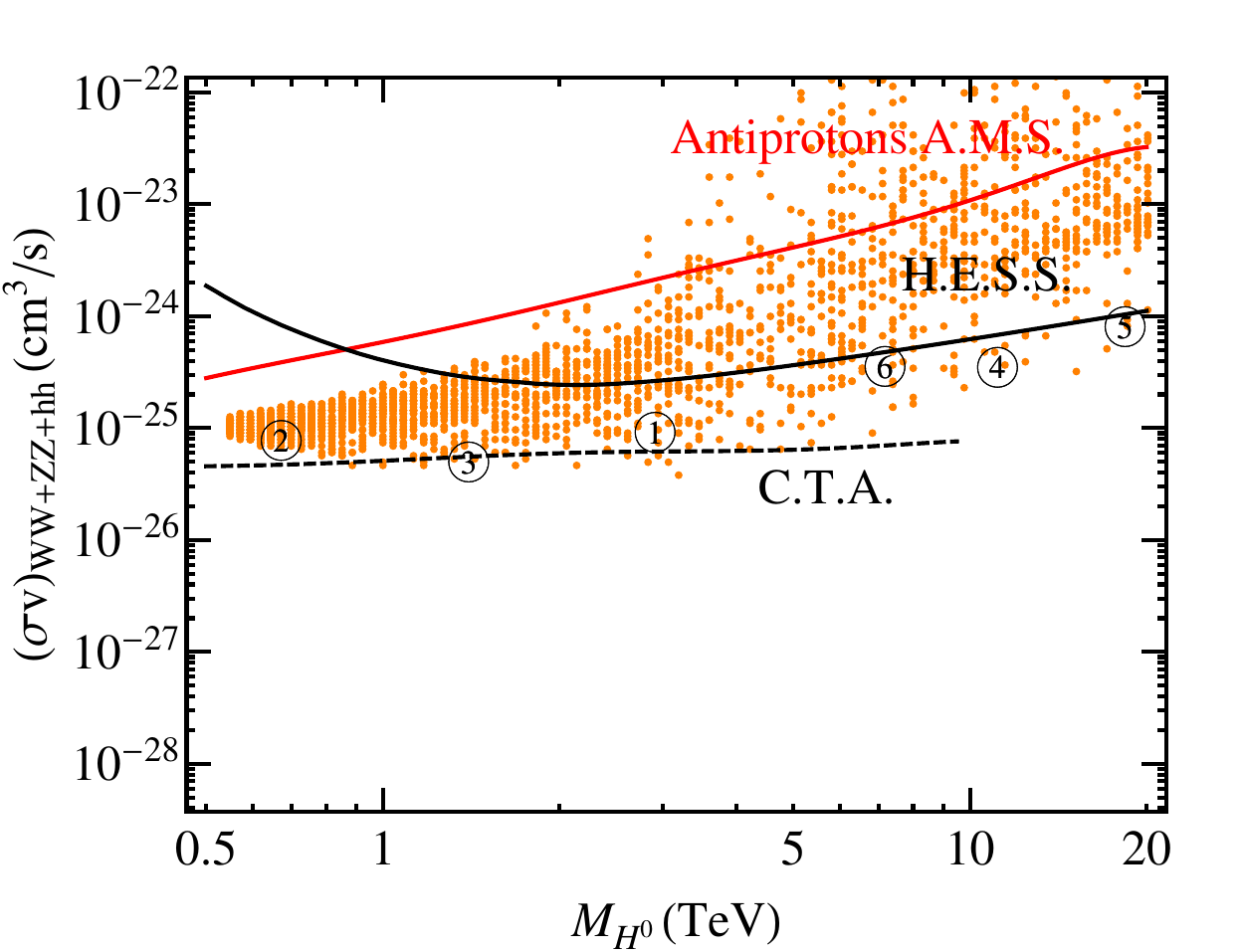}\\
\includegraphics[width=0.49\textwidth]{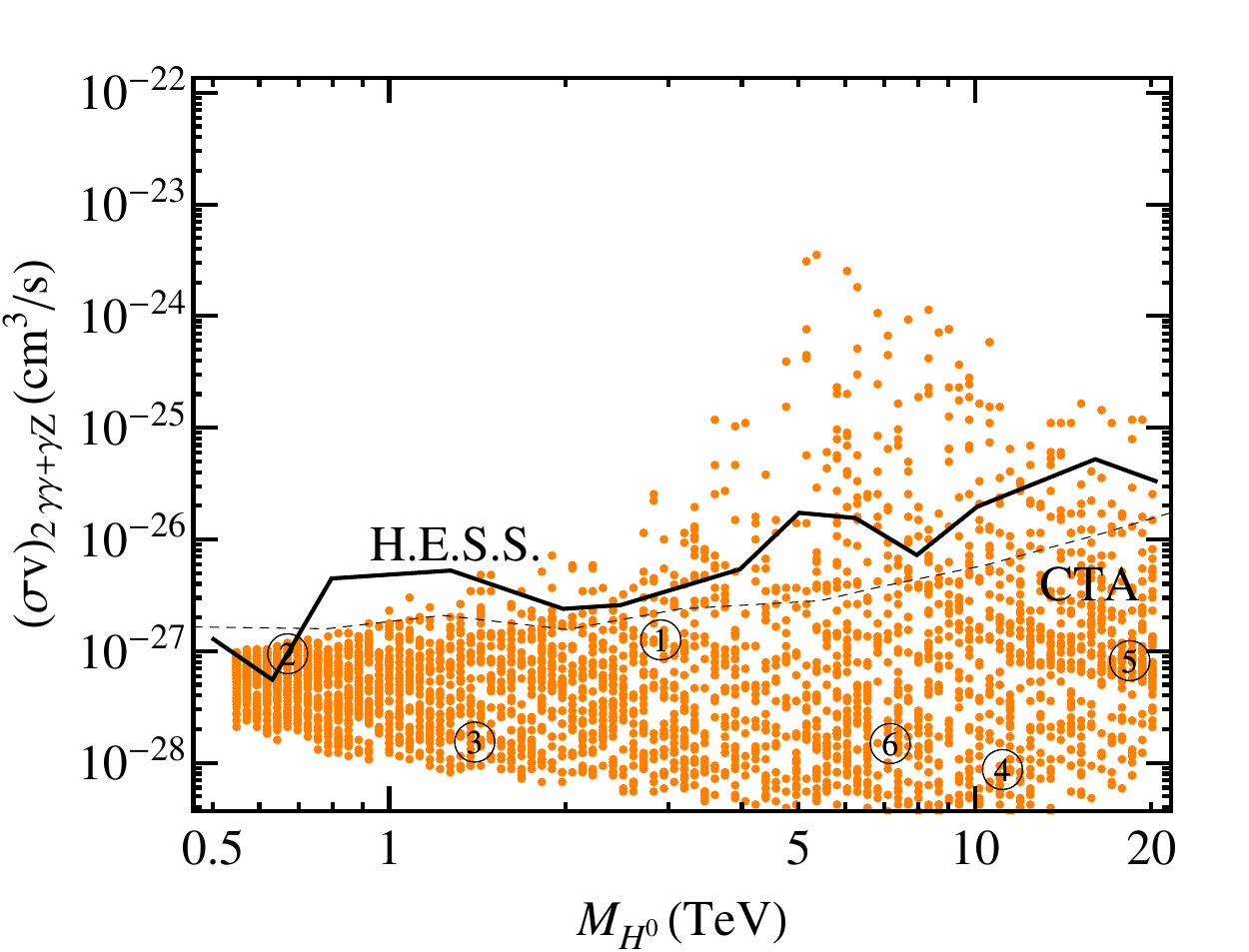}
\includegraphics[width=0.49\textwidth]{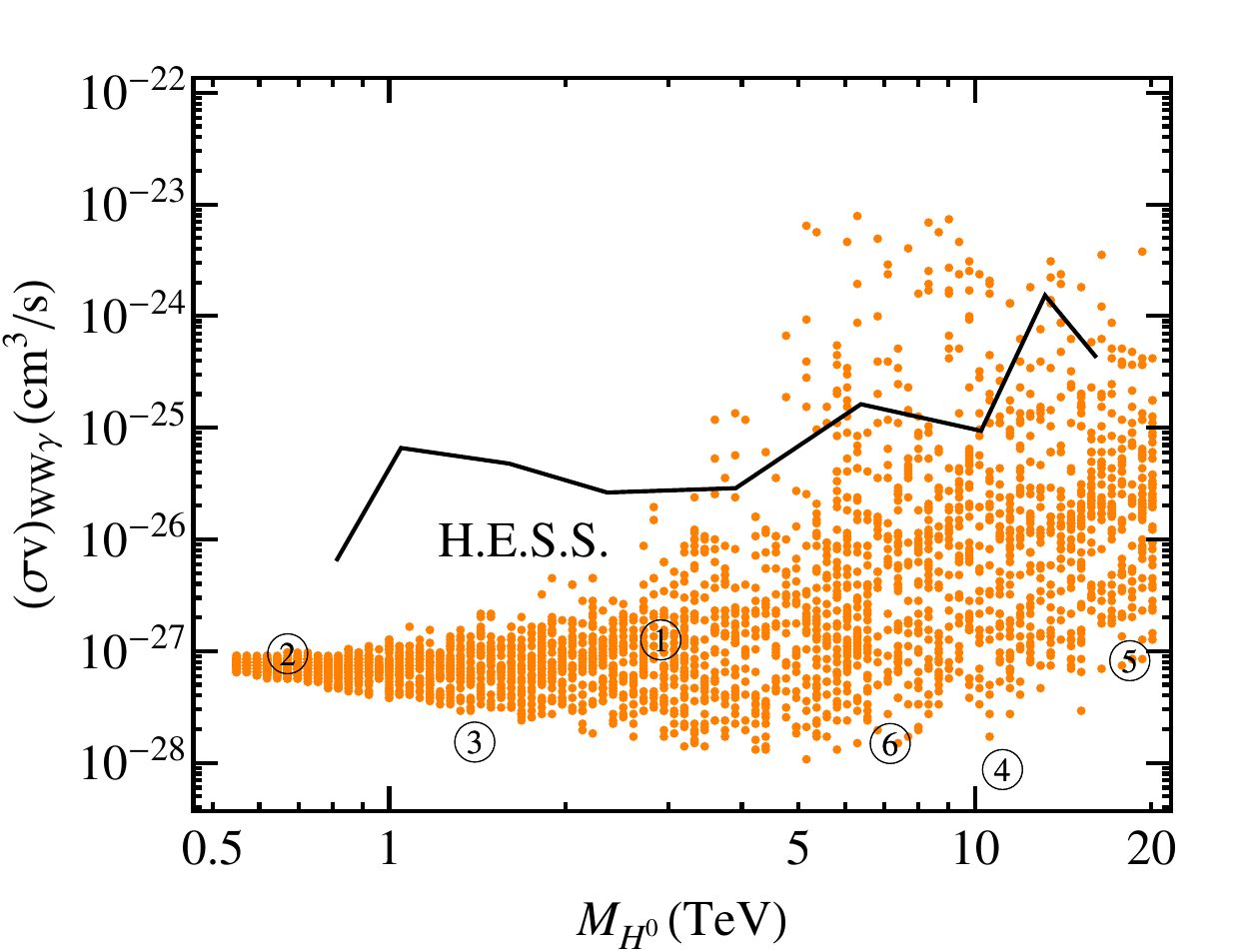}
\caption{\small Annihilation cross section into continuum photons (upper panel), gamma-ray lines (lower-left panel) and virtual internal bremsstrahlung (lower-right panel) for a random sample of points of the viable parameter space (orange points), highlighting the six benchmark points of Table \ref{table:Spectra}, compared to various current upper limits on the cross sections, as well as the projected reach of CTA. For details, see the main text.}
\label{fig:x-sections}
\end{center}
\end{figure}

For all viable points from our scan, we find fairly large annihilation cross sections for both the channels producing  continuum gamma-ray emission and those producing sharp gamma-ray spectral features. These signals could therefore be in reach by present and upcoming gamma-ray telescopes. In Fig.~\ref{fig:x-sections} 
we show in the upper panel the sum of the cross sections of the channels that produce a broad continuum gamma-ray spectrum ($H^0 H^0\rightarrow W^+W^-,~ZZ,~hh$), in the lower-left panel the sum of the channels that produce gamma-ray lines ($H^0 H^0\rightarrow \gamma\gamma,~\gamma Z$)  and in the lower-right panel the channel producing the internal bremsstrahlung signal ($H^0 H^0\rightarrow W^+W^-\gamma$). In the figures, we also highlight our  six benchmark points from Table \ref{table:Spectra} by the tags \circled{1} to \circled{6}. 

The upper plot of Fig.~\ref{fig:x-sections} also includes our derived limits on the cross section into the final state $W^+W^-$  (solid black line) . These limits are calculated by adapting the same procedure as in the H.E.S.S.\ collaboration publication \cite{Abramowski:2011hc} (to be briefly described in the next section). These limits, and all the limits {\it we} derived in this paper, are under the assumption of the Einasto DM density profile using a local DM density of $\rho_\odot = 0.39$~GeV/cm$^3$ and our various data sets are from the inner Galactic center sky region specified in \cite{Abramowski:2011hc,Abramowski:2013ax} unless otherwise stated.
In the same plot we also include  limits from the analysis of Ref.~\cite{Giesen:2015ufa} using preliminary measurements of  the cosmic antiproton-to-proton fraction by the AMS-02 collaboration~\cite{AMS02} (solid red line). 
The two plots in the lower panel instead includes the limits derived by the H.E.S.S. collaboration on the channels $\gamma\gamma$ and $W^+ W^-\gamma$~\cite{Abramowski:2013ax}, respectively.  Furthermore, the Fermi-LAT Collaboration searches for gamma-ray signals from dwarf spheroidal galaxies provide relevant limits \cite{Ackermann:2015zua}. However,  the DM annihilation cross sections predictions can be somewhat different in these galaxies, because the DM velocity dispersion is lower there than in the Galactic center region, and consequently we do not include them in our analysis. 

To examine the expected reach of the upcoming CTA telescope, we show in the upper plot
the projected limits on annihilations into $W^+W^-$ as derived in \cite{Silverwood:2014yza}, assuming 100 hours of observation of a Milky Way center region, and in the lower right plot the limit predictions on monochromatic photons from \cite{Ibarra:2015tya} 
(after a proper rescaling of their limits on narrow boxed shaped spectra), assuming an observation time of  112 hours of the Galactic center region given in \cite{Abramowski:2011hc,Abramowski:2013ax}.

These estimates indicate that present instruments are already sensitive to large regions of the viable parameter space of the IDM and that CTA has good prospects to observe a signals from this model by the observation of a broader continuum excess in the gamma-rays spectrum. Furthermore, the continuum signal flux might be complemented by a simultaneous univocal DM signal in the form of a sharp gamma-ray spectral feature. 

\section{Complementarity among Gamma-ray Signals in the IDM}
\label{sec:FvsC}

In order to more carefully asses the prospects to observe signals of the IDM, we  derive dedicated limits on each of the IDMs from our parameter scan. For each model's cross section limit, we then define a maximal boost factor (BF) that corresponds to how much the model's predicted gamma-ray signal can be increased before it saturates its derived limit.

\begin{figure}[t!]
\begin{center}
\includegraphics[width=0.49\textwidth]{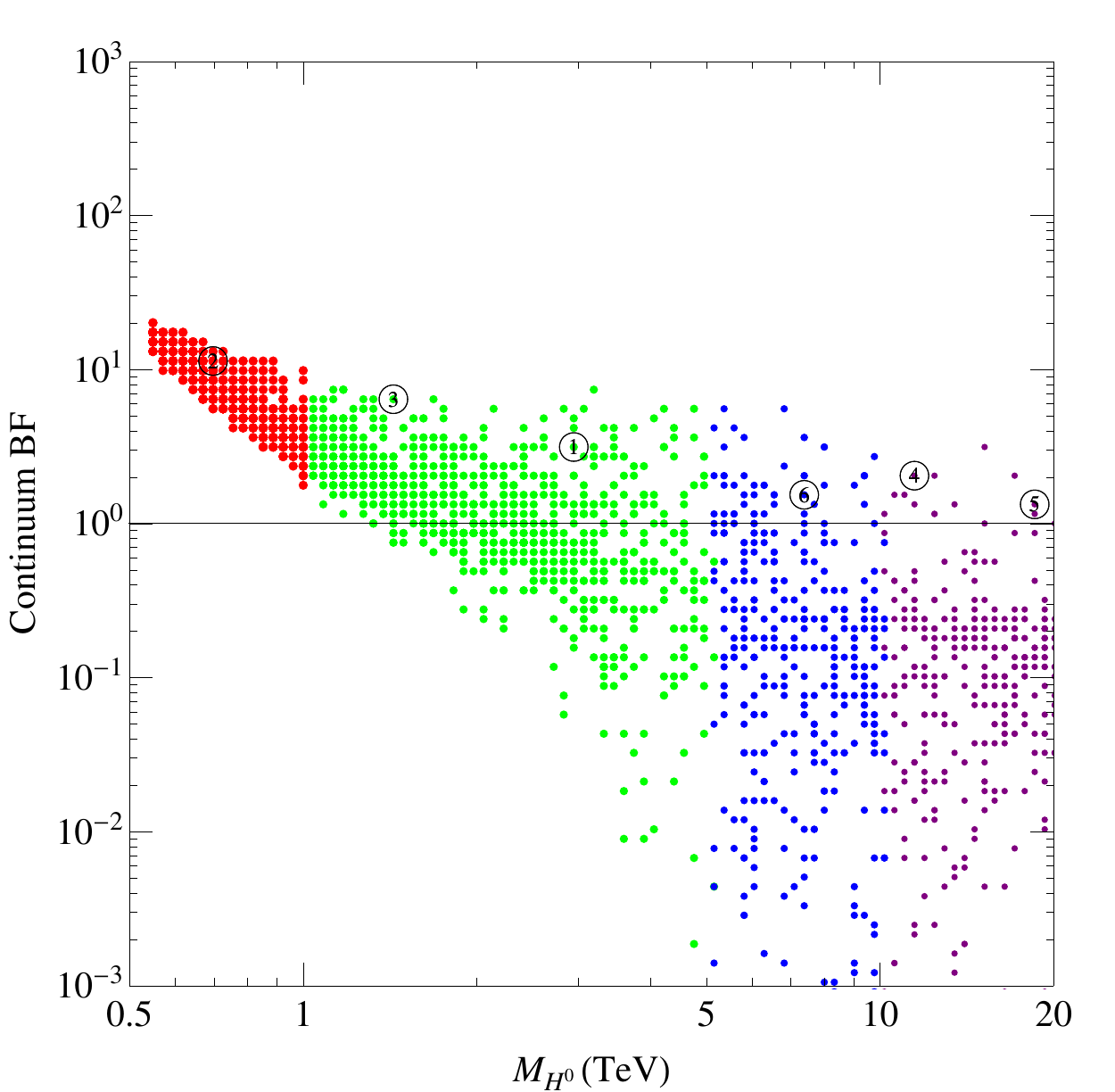}
\includegraphics[width=0.49\textwidth]{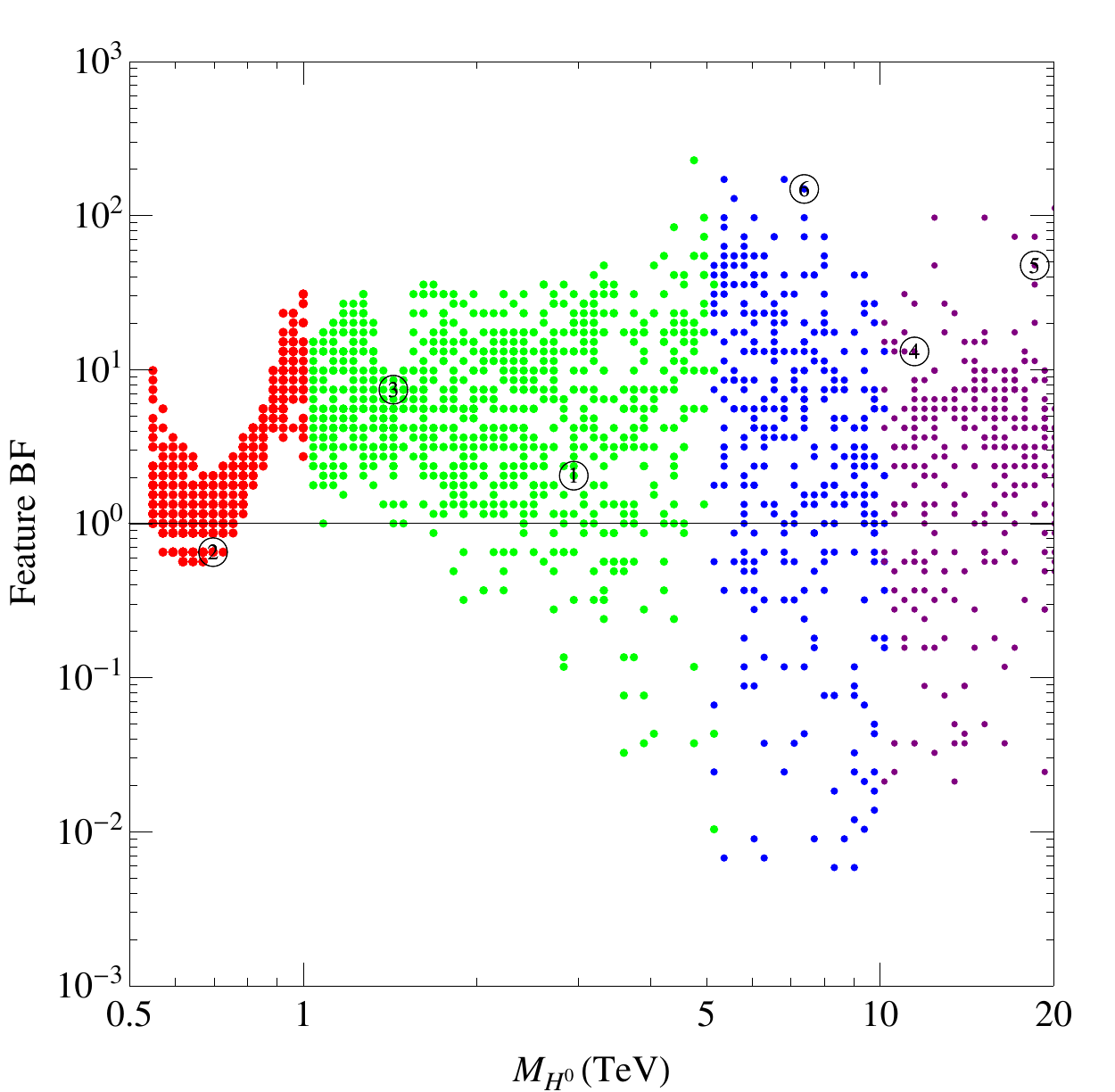}
\caption{\small Upper limit on the boost factor from the non-observation of the continuum part of the gamma-ray spectrum (left plot) and of the sharp spectral features (right plot) expected from annihilations in the IDM for a random sample of points of the viable parameter space, highlighting the  six benchmark points  of Table \ref{table:Spectra}. The color of the points indicate the mass range where they fall. For details, see the main text. 
}
\label{fig:boost}
\end{center}
\end{figure}

To derive signal limits on each IDM induced continuum signal,  we  use the data collected by the H.E.S.S.\ instrument and closely follow the method pursued in \cite{Abramowski:2011hc}, which compares the  gamma-ray fluxes measured  in a ``search region'' and in a ``background region''  around the Galactic center. The $\bar{J}$-factors in the search and background regions are given, respectively, by $\bar{J}= 7.41 \times 10^{24}\,\text{GeV}^2\,\text{cm}^{-5}$ and $\bar{J}= 3.79 \times 10^{24}\,\text{GeV}^2\,\text{cm}^{-5}$~\cite{Abramowski:2011hc}. This is the same procedure we used to derive the  $W^+W^-$ limits from H.E.S.S.\ for Fig.~\ref{fig:x-sections} in the previous section. The derived BFs from the IDMs induced continuum gamma-ray signals are shown in the left plot of Fig.~\ref{fig:boost}.
Points with DM mass in the range $0.5\,\text{TeV}< M_{H^0}< 1\,\text{TeV}$, $1\,\text{TeV}< M_{H^0}< 5\,\text{TeV}$, $5\,\text{TeV}< M_{H^0}< 10\,\text{TeV}$ and $10\,\text{TeV}< M_{H^0}< 20\,\text{TeV}$ are shown in the colors (to be used also for future references) red, green, cyan and blue, respectively.  Among the viable models, the six benchmark points  of Table \ref{table:Spectra} are highlighted in the plot with their corresponding tag. 

Notably, there are many points, especially with mass above $\sim 2$~TeV which are already excluded by observations with the H.E.S.S. instrument. Furthermore, for most of the points the BF value is constrained to be smaller than $\sim 10$. Therefore, an improvement in sensitivity of gamma-ray telescopes to this type of exotic continuum flux by a factor of $\sim 10$, which seems to be feasible with the upcoming CTA (see {\it e.g.} Ref.~\cite{Silverwood:2014yza} and our Fig.~\ref{fig:x-sections}), could suffice to cover all the signal predictions from the IDM, assuming that the DM halo distribution follows the Einasto profile. For a Navarro-Frenk-White profile of the DM distribution, the $\bar{J}$ factor in the target region is a factor of  two smaller \cite{Abramowski:2011hc}, hence the annihilation signal would in this case be a factor of two fainter and the prospects for detection, somewhat poorer. 

We have also calculated the maximal boost factor BF from the non-observation of the sharp gamma-ray spectral features produced by the final states $\gamma\gamma$, $\gamma Z$ and $W^+W^-\gamma$. These limits on the IDM were derived following the procedure pursued  by the H.E.S.S. collaboration in \cite{Abramowski:2013ax}, which adopts a phenomenological background model defined by seven parameters. The result of this procedure is illustrated in Table~\ref{table:Fluxes} for the four first benchmark points  of Table \ref{table:Spectra}, which are IDMs characterized by having an intense sharp gamma-ray spectral feature. 
\begin{table}[t]
\centering
\begin{tabular}{|c|c|} \hline
\multicolumn{2}{|c|}{\bf Benchmarks} 
\\\hline
%
BMP 1 
& BMP 2\\
\includegraphics[trim=0cm 0cm 0cm 0.8cm,clip,width=7.4cm]{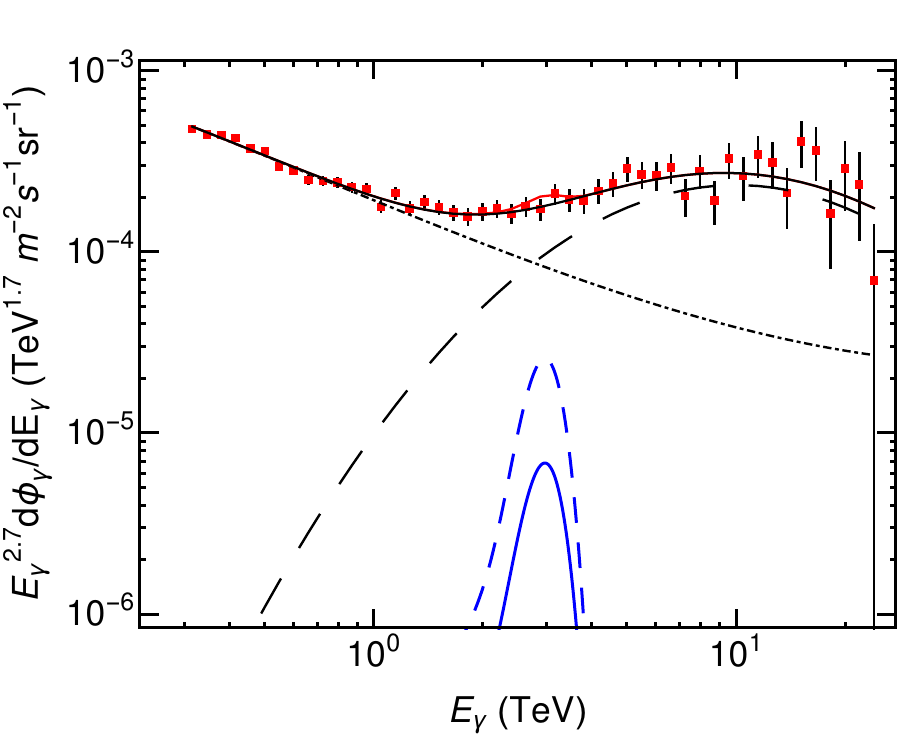}
&
\includegraphics[trim=0cm 0cm 0cm 0.8cm,clip,width=7.4cm]{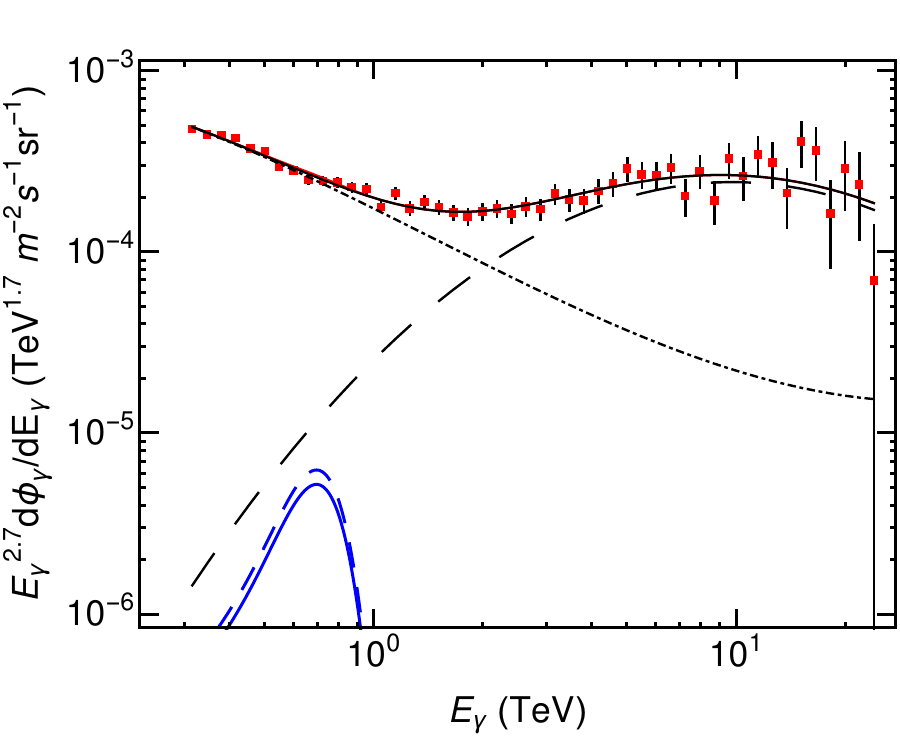}
\\\hline
%
BMP 3 
& BMP 4\\
\includegraphics[trim=0cm 0cm 0cm 0.8cm,clip,width=7.4cm]{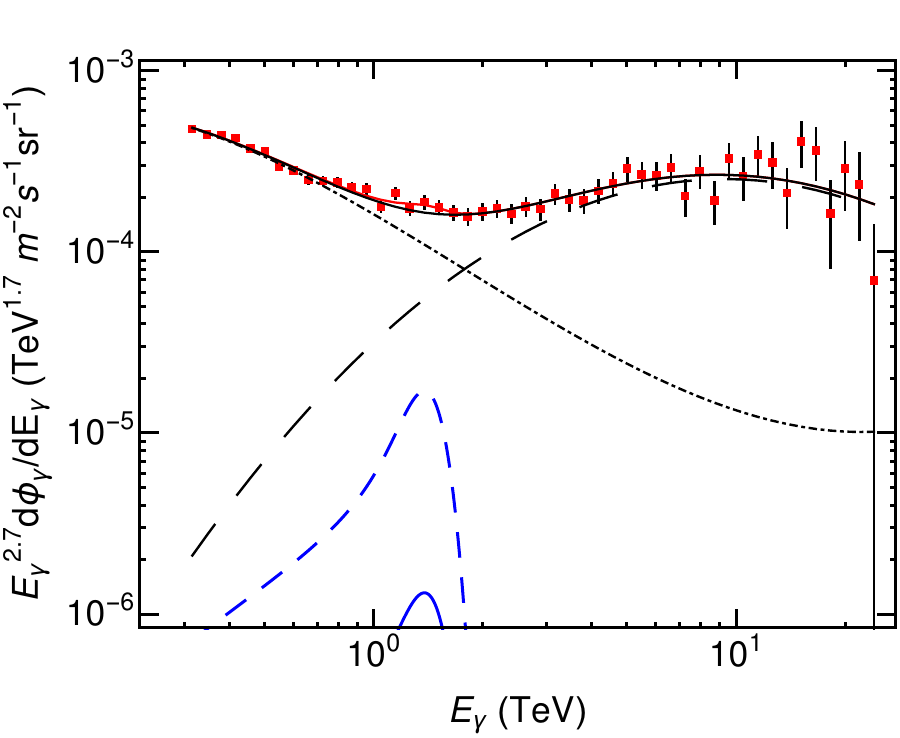}
&
\includegraphics[trim=0cm 0cm 0cm 0.8cm,clip,width=7.4cm]{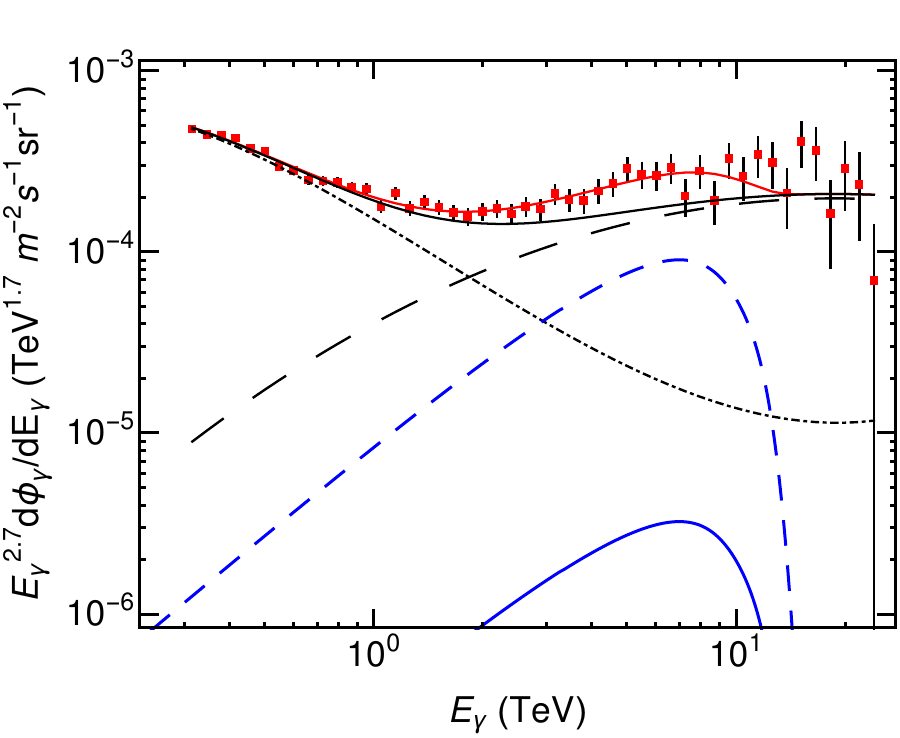}
\\\hline
\end{tabular}
\caption{\small H.E.S.S. limits on the benchmarks of Table \ref{table:Spectra}. See the text for details.}
\label{table:Fluxes}
\end{table}
%
In the figures of Table~\ref{table:Fluxes}, we show the predicted DM signal (solid blue line) and the DM signal after being enhanced by the boost factor BF (dashed blue line) constructed to saturate the derived 95\% C.L.\ limits from the H.E.S.S.\ data (shown by the red dots) \cite{Abramowski:2013ax}. The best-fit background model for the BF enhanced signal is shown by the solid black line and the total gamma-ray flux model, including the enhanced DM signal, are shown by the solid red line.
The boost factor BF for our sample of points derived from the non-observation of a sharp spectral feature with H.E.S.S.\ is shown  in the right panel of Fig.~\ref{fig:boost}. We find again models which are already ruled out by present observations, especially at large DM masses.\footnote{The strengthening of the limits at $M_{H^0}\sim 600$ GeV is due to a dip around $E_\gamma\sim 700$ GeV in the gamma-ray flux measured by the H.E.S.S. collaboration, and which is possibly due to a downward statistical fluctuation.} Furthermore, with an increase in sensitivity by a factor $\sim 10$, which is likely to be achieved with the upcoming CTA (see {\it e.g.} \cite{Ibarra:2015tya, Garcia-Cely:2015dda}), a significant part of the IDM parameter space will be probed, thus opening the exciting possibility of observing unambiguous signals from DM annihilation at future gamma-ray telescopes. Unfortunately, to guarantee the observation of a sharp feature in the gamma-ray spectrum a larger increase in sensitivity is necessary, concretely by a factor $\sim 100$, assuming the Einasto profile.

\begin{figure}[t]
\begin{center}
\includegraphics[width=0.7\textwidth]{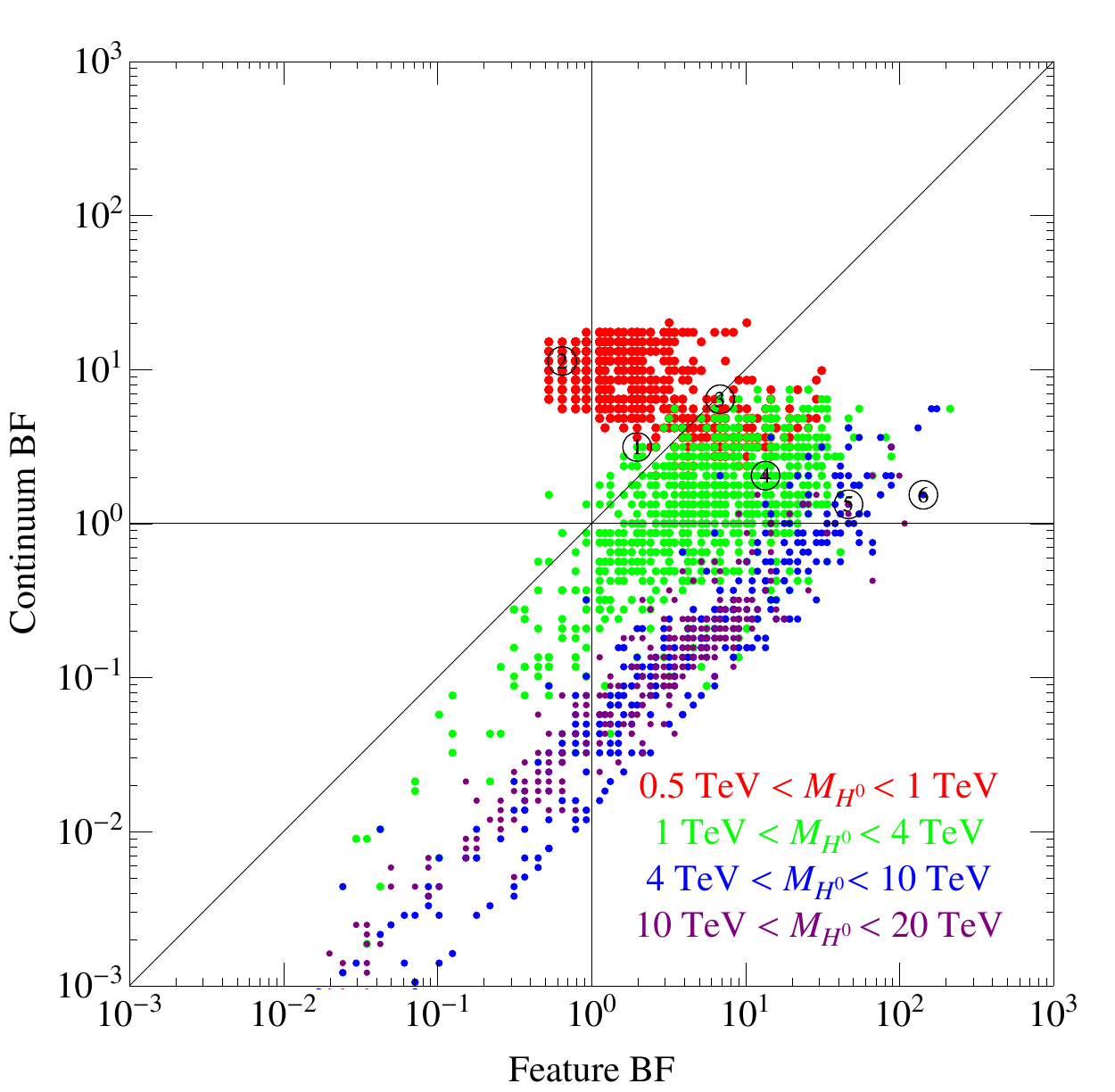}
\caption{Complementarity between the upper limits on the boost factor from the non-observation the continuum part of the gamma-ray spectrum and from the non-observation of sharp gamma-ray spectral features. The color and tagging of the points is as in Fig.~\ref{fig:boost}.}
\label{fig:cont_vs_line}
\end{center}
\end{figure}

From the above discussions it apparently becomes relevant to investigate the potential complementarity between the searches for a continuum exotic flux and a sharp spectral feature. In Fig.~\ref{fig:cont_vs_line}  
we illustrate this complementary. For each model in our scan, the required boost factor for a model to become excluded by the conti-nuum spectrum constraints (continuum BF) is confronted to the required boost factor to become excluded by the sharp spectral feature (feature BF). It follows from the figure that, for most of the points, the former provides a stronger constraint than the latter, with the difference between the BFs being more notable as the DM mass increases. This behaviour is a consequence of the requirement on the  parameters to correctly reproduce the DM density via thermal freeze-out. As argued in Section~\ref{sec:DMHiggs}, larger and larger quartic couplings are required when the DM mass increases. As a result, the annihilation rates  into $W^+W^-$, $ZZ$ and $hh$, which  can be induced by quartic coupling interactions (as follows from Eqs.~\eqref{GammaZZ}, \eqref{GammaWW} and \eqref{Gammahh}),  are enhanced compared to $\gamma\gamma$ and $\gamma Z$, which are induced only by gauge interactions (as follows from  Eq.\eqref{Gammagamma}).

\bigskip
We would like to briefly comment on the difference between the IDM and  two other minimal DM scenarios with a DM candidate in a 5-plet and 7-plet representations of $SU(2)_L$ \cite{Garcia-Cely:2015dda,Cirelli:2015bda, Aoki:2015nza}. In the these papers it was concluded that if the DM candidates account for all the DM then they are excluded if their masses are below 20\,TeV.  Although we cover similar masses in the IDM, not all them are excluded, as it is shown in Fig.~\ref{fig:cont_vs_line}. The underlying reason for this is related to the mass splittings between the charged and the neutral components of the inert states. In the minimal 5-plet and 7-plet  DM scenarios, the mass splitting is set by radiative corrections and is fixed to a constant value. In the IDM there is more freedom, and the mass splitting can be larger than the quantum effect. In fact, as already discussed, for large DM masses, large quartic couplings are  typically needed to achieve the right relic abundance. Unless a cancellation between $\lambda_4$ and $\lambda_5$ takes place, the mass difference between $H^+$ and $H^0$ becomes relatively larger. This leads to relatively smaller Sommerfeld effects in comparison to these minimal DM models (even if it is still large for the heaviest DM masses in IDM). Another important reason is that larger $SU(2)_L$ multiplets will contain particles with  larger electric charges. This leads to larger annihilation cross sections for all gauge mediated annihilation channels and, in particular, increases monochromatic gamma-ray signals.

\section{Complementarity with Direct Detection} 
\label{sec:DD}

\begin{figure}[t!]
\begin{center}
\includegraphics[width=0.7\textwidth]{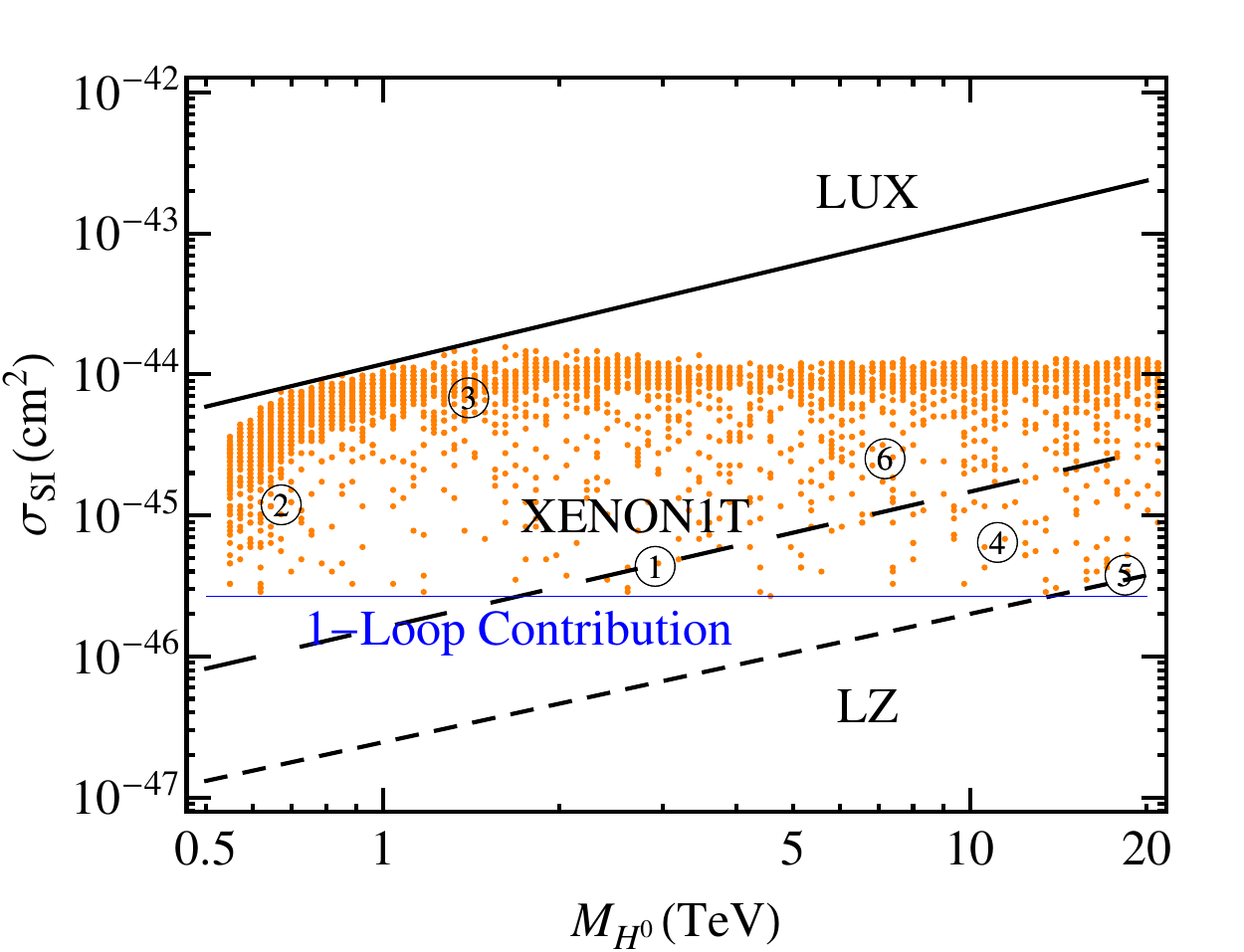}
\caption{Predicted value of the spin-independent DM scattering cross section with protons for a random sample of viable points of the IDM (orange lines), highlighting the six benchmark points of Table \ref{table:Spectra}, compared to the current upper limit from the LUX experiment (solid black line) as well as the projected sensitivities of the XENON1T (long dashed) and LZ (short dashed) experiments. The {minimal} value of the cross section induced by the one-loop exchange of weak gauge bosons is shown as a solid blue line. 
}
\label{fig:SI}
\end{center}
\end{figure}

A complementary avenue to probe the high mass regime of the IDM is direct detection. The spin-independent scattering cross section of DM particles with nuclei receives in this model two different contributions. The first one is induced by the  t-channel exchange of a Higgs boson  with the nucleon $n$~\cite{Barbieri:2006dq}:
\beq
\sigma_{\text{SI}}=\frac{M_n^4 (\lambda_3+\lambda_4+\lambda_5)^2f^2}{
{4}\pi(M_n+M_{H^0})^2M_h^4},
\label{eq:SSI}
\eeq
where $f\approx 0.3$ is a form factor with its precise value taken from micrOMEGAs 3.1~\cite{Belanger:2013oya} and $M_n = 0.939$~GeV is the nucleon mass. This cross section is suppressed for large DM masses and for small DM-Higgs coupling, which corresponds to $|\lambda_3+\lambda_4+\lambda_5|$. The second contribution is induced by one-loop exchange of gauge bosons~\cite{Cirelli:2005uq, Hambye:2009pw, Klasen:2013btp}, which is independent of the quartic couplings and which sets a lower limit on the interaction cross section of $\sigma_\text{SI}\gtrsim 2.6\times 10^{-46}\,\text{cm}^2$ independently on the DM mass. We show in Fig.~\ref{fig:SI} 
the predicted spin-independent scattering cross section with protons for our sample of viable points and compare them to the limit from the LUX experiment as well as to the projected reach of XENON1T~\cite{Aprile:2012zx} and LZ~\cite{Akerib:2015cja}. As apparent from the plot, the current data from the  LUX experiment barely constrain the viable parameter space of the IDM. On the other hand, and due to the above-mentioned lower limit on the interaction cross section induced by the one-loop exchange of gauge bosons, the upcoming XENON1T (LZ) experiment should observe a signal of the IDM for thermally produced DM particles with $M_{H^0}\lesssim 1.6\,\text{TeV}~(13\,\text{TeV})$. We note that many of the points of our scan have an interaction cross section $\sigma_\text{SI}\gtrsim 5\times 10^{-45}\,\text{cm}^2$, which is well within the reach of XENON1T.

\begin{figure}[t!]
\begin{center}
\includegraphics[width=0.49\textwidth]{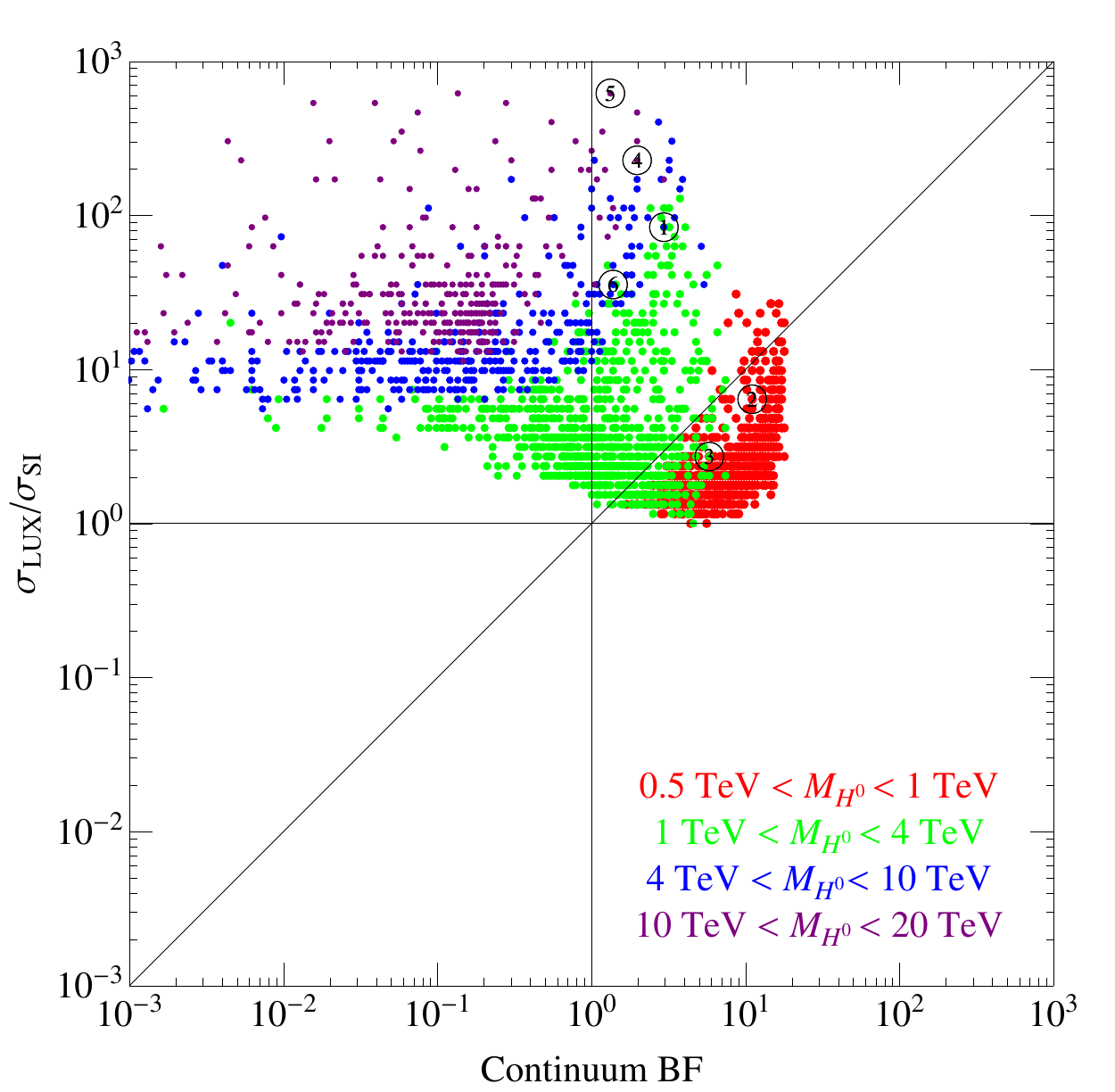}
\includegraphics[width=0.49\textwidth]{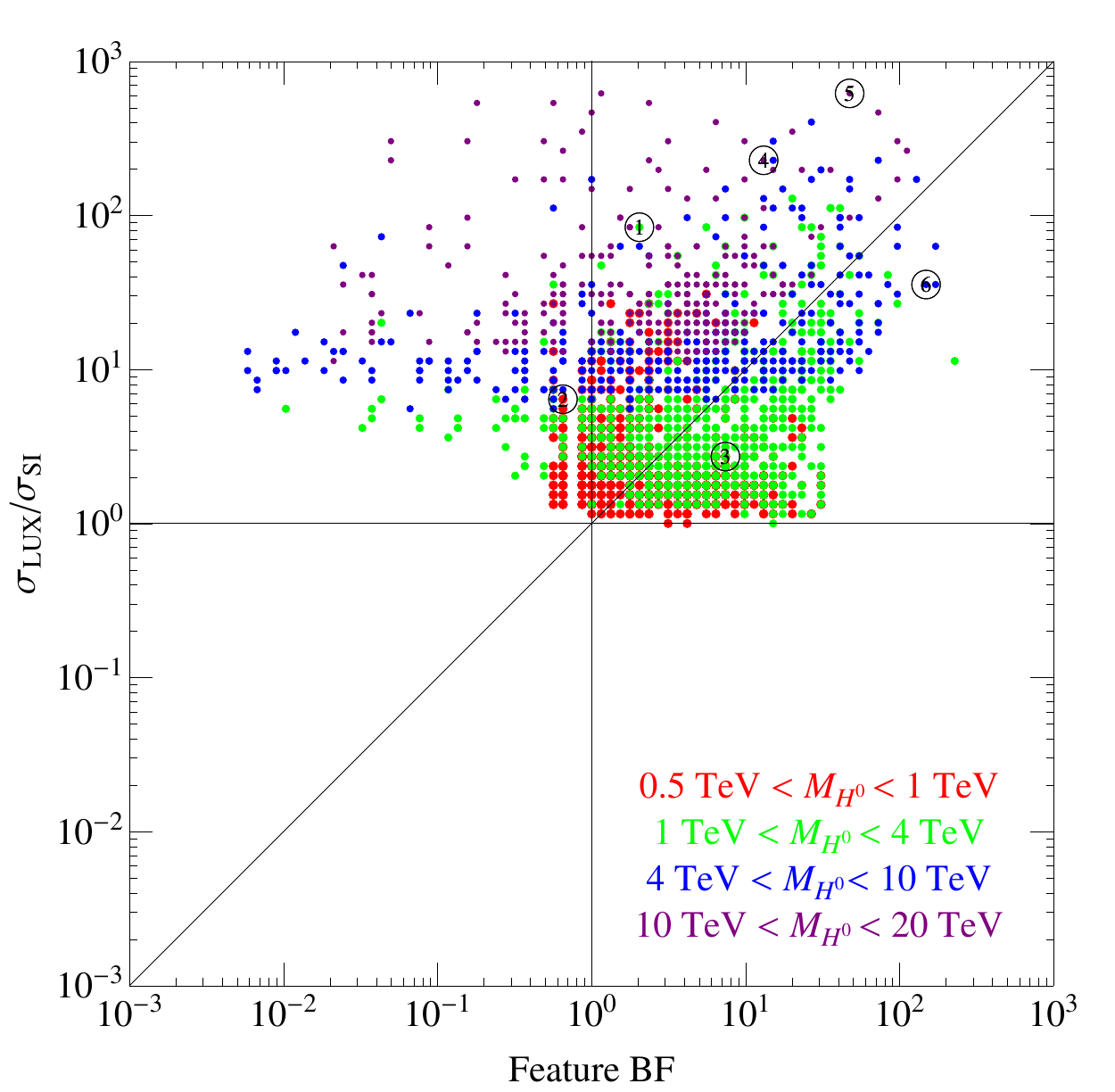}
\caption{Complementarity between the direct detection limits on the IDM from LUX, and the upper limits on the boost factor from the non-observation the continuum part of the gamma-ray spectrum (left panel) and from the non-observation of sharp gamma-ray spectral features (right panel). The color and tagging of the points is as in Fig.~\ref{fig:boost}.}
\label{fig:DD_vs_gamma}
\end{center}
\end{figure}

We finally confront the constraints from direct detection DM searches to those from indirect detection searches in gamma-rays. In Fig.~\ref{fig:DD_vs_gamma},
the left plot (right plot) shows the viable DM models in the plane of the ratio $\sigma_\text{LUX}/\sigma$ against the required boost factor of the broad DM gamma-ray signal  (sharp spectral feature) to reach the current H.E.S.S.\ constraints. Here $\sigma$ is the predicted spin-independent cross section with a proton and $\sigma_\text{LUX}$ is the upper limit from LUX  for that model.
Points with $\sigma_\text{LUX}/\sigma<1$ are then ruled out by the LUX experiment and points with BF$<1$  are ruled out by the H.E.S.S. instrument. The values of BF and $\sigma_\text{LUX}/\sigma$ for our six benchmark points are indicated in the figure and displayed  for reference in Table \ref{table:LimitsI}.
\begin{table}[t]
\centering
\begin{tabular}{|c|c|c|c|c|c|}\hline
BMP  & Boost Feature  & Boost Continuum & $\sigma_\text{LUX}/\sigma_\text{SI}$\\\hline\hline
1 & 1.93 & 2.87  & 77.2 \\\hline
2 & 0.63 & 10.76 & 6.3  \\\hline
3 & 6.67 & 5.71 & 2.4  \\\hline
4 & 13.01 & 1.95 & 204.0 \\\hline
5 & 44.9 & 1.29 & 574.1  \\\hline
6 & 141.0 & 1.34 & 33.5  \\\hline
\end{tabular}
\caption{\small Upper limit on the feature boost factor and continuum boost factor from current gamma-ray telescopes, as well as value of the spin-independent cross section relative to the current upper limit from LUX, for the six benchmark points defined in table~\ref{table:LimitsI}. }
\label{table:LimitsI}
\end{table}
It follows from the figures that, for low DM masses, namely $M_{H^0}\lesssim 5$ TeV, an  increase of sensitivity in direct detection experiments and gamma-ray telescopes by a factor 10, which seems feasible in the near future, might lead to three different DM signals. An exciting possibility is thus to detect: a scattering signal in a direct detection experiment, a broad gamma-ray excess from the central galactic halo region together with a  sharp spectral feature in the gamma-ray spectrum. On the other hand, for larger DM masses, indirect detection with gamma-rays constitutes the most promising search strategy for the IDM.

\section{Conclusions}

\label{sec:conclusions}

We have investigated the gamma-ray spectrum produced in DM annihilations in the center of the galaxy for the high mass regime of the Inert Doublet Model (IDM). We have found that,  in order to satisfy the requirements of unitarity on the annihilation cross sections, it is necessary to account for  the so-called Sommerfeld enhancement. This is a non-perturbative effect arising from the exchange of  gauge and  Higgs bosons between non-relativistic annihilating DM particles. In the mass regime under consideration, such exchange induces long range interactions that lead to a significant modification of the annihilation cross sections. We have argued that including that effect is crucial for phenomenological studies of indirect DM signals from the galactic center, specially for masses much larger than 1\,TeV and small mass splittings between the charged and the neutral scalars. We have also showed that such effect is much less important for the DM production in the early Universe (see Appendix~\ref{sec:AppendixC}).

We have calculated the impact of the Sommerfeld enhancement in the framework of the effective field theory resulting from the non-relativistic limit of the inert scalar particles. The main ingredients to consider are the potential matrices of Eq.~\eqref{potential}, encoding the long-range effects, and the  matrices of Eqs.~\eqref{Gammagamma}, \eqref{GammaWW}, \eqref{GammaZZ} and \eqref{Gammahh}, which describe the annihilation processes. In appendix~\ref{sec:AppendixB} we have described succinctly how to use these matrices in order to the obtain the annihilation cross sections.  Using this formalism, we have been able for the first time to reliably calculate the gamma-ray spectrum. It receives contributions from (i) a featureless soft part arising from annihilations into $W^+W^-$, $ZZ$ and $hh$ pairs, (ii) monochromatic photons in the $\gamma\gamma$ and $\gamma Z$ final states and (iii) from the virtual internal bremsstrahlung process $H_0 H_0 \to W^+W^-\gamma$. In table \ref{table:Spectra},  we have presented a set of benchmark points (BMP1$-$BMP6), compatible with all theoretical and experimental constraints  on the IDM (see Appendix~\ref{sec:AppendixA}) and have classified them according to the relative importance of each contribution. BMP1 and BMP2 exhibit strong spectral features at the end point of their spectra, BMP3 and BMP4 receive their most important contribution from virtual internal bremsstrahlung, and BMP5 and BMP6 are dominated by the broader continuum emission of photons.  

We have then confronted the IDM with the most recent Galactic center observations of the H.E.S.S. instrument.  Assuming the Einasto profile, and using a scan over the five-dimensional parameter space of the DM particle model sector, we have found that many viable models are already excluded by H.E.S.S., mostly via the continuum emission. The result of this is shown in Fig.~\ref{fig:cont_vs_line}. We would like to remark that  H.E.S.S. can probe many viable DM models even though it does not currently reach annihilation cross sections close to the canonical  thermal value of $3\times 10^{-26}$cm$^3$/s. The underlying reason for this is the Sommerfeld enhancement and, to a lesser extent,  coannihilations in the early Universe~\cite{Queiroz:2015utg}.

Subsequently, we have analyzed the interplay between this indirect search and the  direct DM searches with the LUX experiment. The corresponding results are shown in Fig.~\ref{fig:DD_vs_gamma}, and for our benchmark points, on table~\ref{table:LimitsI}.  Current direct DM searches are not sensitive enough to detect signals of the IDM. Nevertheless, the upcoming XENON1T experiment and the projected LZ experiment will be able to close in on the viable parameter space of the model. 

Finally, we would like to comment on the sensitivity of the upcoming  Cerenkov Telescope Array to the IDM. As shown in Fig.~\ref{fig:x-sections}, a significant part of the viable models of our scan can be potentially probed by CTA  since most of them produce a continuum spectrum that is within a factor ten from current experimental sensitivity.

\bigskip
\textbf{Note Added}\\

During the last stages of this work, we learned of the analysis of Ref.~\cite{Queiroz:2015utg}, where  the gamma-ray signals of the IDM in its high mass regime are also studied. In that paper, however, the Sommerfeld effect was neglected and the contribution to the spectrum from gamma-ray lines and from virtual internal bremsstrahlung was not considered.

\acknowledgments{
The authors would like to thank Anna S.~Lamperstorfer for useful discussions.  The work of C.G.C. is supported by the FNRS, the IISN and the 
Belgian Federal Science Policy through the Interuniversity Attraction Pole P7/37. M.G.\ acknowledges partial support from the European Union FP7 ITN Invisibles (Marie Curie Actions, PITN-GA-2011-289442).  The work of AI was partially supported by the DFG cluster of excellence ``Origin and Structure of the Universe''.}

\appendix
\section{Constraints on the Model Parameters}
\label{sec:AppendixA}

We summarize here the various theoretical and phenomenological constraints on the scalar potential which are included in our scan of the parameter space of the IDM.
\begin{itemize}
\item{\bf Perturbativity} can, at least naively, be enforced by assuming that no scalar coupling  exceeds $4\pi$ (see table~3.2 of \cite{GarciaCely:2014jha})
\begin{align}
| \lambda_{1,2,3,4,5}| \leq 4\pi \,,\hspace{10pt}
|\lambda_3+\lambda_4\pm\lambda_5| < 4\pi \,,\hspace{10pt}
|\lambda_4\pm\lambda_5| < 8\pi \,,\hspace{10pt}
|\lambda_3+\lambda_4| < 4\pi. 
\end{align}
\item{\bf Vacuum stability} of the potential requires \cite{Gunion:2002zf,Gustafsson:2010zz} (see also \cite{Khan:2015ipa})
\begin{equation}
\lambda_1 > 0\;, \hspace{10pt}\lambda_2 > 0\;, \hspace{10pt}\lambda_3 > -2 (\lambda_1\lambda_2)^{\frac{1}{2}}\;, \hspace{10pt}\lambda_3+\lambda_4-|\lambda_5| > -2 (\lambda_1\lambda_2)^{\frac{1}{2}}.
\label{treelevelconstraint}
\end{equation}

\item{\bf Unitarity of the S-matrix} on scalar to scalar, gauge boson to gauge boson and scalar to gauge boson scatterings at the perturbative level furthermore requires that~\cite{Ginzburg:2004vp,Branco:2011iw} 
\begin{align}
 \lambda_3 \pm \lambda_4                                                                  		\leq  8 \pi, \quad
\lambda_3 \pm \lambda_5                                                                 		\leq   8 \pi, \quad
 \lambda_3+ 2 \lambda_4 \pm 3\lambda_5                                                        &\leq  8 \pi\nn\\
 -\lambda_1 - \lambda_2 \pm \sqrt{(\lambda_1 - \lambda_2)^2 + \lambda_4^2}                        		& \leq  8 \pi\nn\\
 -3\lambda_1 - 3\lambda_2 \pm \sqrt{9(\lambda_1 - \lambda_2)^2 + (2\lambda_3 + \lambda_4)^2}  & \leq  8 \pi\nn\\
 -\lambda_1 - \lambda_2 \pm \sqrt{(\lambda_1 - \lambda_2)^2 + \lambda_5^2}                      		&\leq  8 \pi.\nn
\label{eq:unitary}
\end{align}
\end{itemize}

\begin{itemize}

\item{\bf Electroweak precision observables} constraints the contributions to the Peskin-Takeuchi $S,\;T,\;U$ parameters to remain in the region
$\Delta S = 0.06\pm 2\times0.09$, $\Delta T = 0.1\pm 2\times0.07$ with a correlation coefficient of +0.091 (when $\Delta U$ is fixed to zero, which is appropriate for the IDM). The contribution from the IDM can be calculated as in, e.g., \cite{Arhrib:2012ia}. This typically prohibit large mass splittings among inert states, but for DM masses with $M_{H^0}\gtrsim 500$~GeV relatively small  splittings are already required, especially when combined with the relic density constraint \cite{Hambye:2009pw}.
\end{itemize}

Besides these theoretical constraints on the parameters of the IDM, there are also limits from experimental searches of the inert scalars. 
\begin{itemize}
\item{\bf LEP bound} comes first from that the decay channels $Z\rightarrow A^0 H^0$, $Z\rightarrow H^+ H^-$, $W^\pm\rightarrow A^0 H^\pm$ and $W^\pm\rightarrow H^0 H^\pm$ would alter the gauge bosons measured mass widths. As a good approximations, these implies that $M_{A^0} + M_{H^0} \geq M_{Z}$, $2M_{H^\pm}\geq M_{Z}$, $M_{H^\pm} + M_{H^0,A^0} \geq M_{W}$. Second, constraints on IDM parameters have been extracted from chargino searches at LEP\,II: The charged Higgs mass is constrained by $M_{H^\pm}\gtrsim 70$~GeV ~\cite{Pierce:2007ut}. The bound on $M_{H^0}$ is more involved: If  $M_{H^0}<80$~GeV then $M_{A^0}-M_{H^0}$ should be less than $\sim 8$ GeV, or else, $M_{A^0}$ should be greater than $\sim 110$~GeV \cite{Lundstrom:2008ai}. 

\item{\bf LHC bounds} come from the Higgs sector. The new scalar states can either increase the invisible branching ratio and/or alter the diphoton signal strength of the Higgs boson \cite{Arhrib:2012ia,Krawczyk:2013jta,Swiezewska:2012eh,Goudelis:2013uca}. These bounds are typically only relevant for masses below $M_h/2$, and thus play a little role for inert scalar particle masses well above. Direct di-lepton searches have also been shown to restrict the inert scalar masses in the region of $M_{H^0}\lesssim 60$~GeV and  $M_A \lesssim 150$~GeV \cite{Belanger:2015kga}. 
\end{itemize}
From these constraints it is clear that the IDM is strongly restricted if the new states are at sub 100 GeV masses and not so constrained for masses above 500\,GeV. In addition there are various astrophysical constraints (see, e.g., the review in \cite{Gustafsson:2010zz}), and direct detection constraints. Due to the relevance of the latter for this work, they are  discussed in the main text.

\section{Algorithm for the Sommerfeld Enhancement}
\label{sec:AppendixB}

In this appendix we briefly describe the algorithm to calculate the Sommerfeld enhancement factors, given the potential of Eq.~\eqref{potential} and the annihilation matrices of  Eqs.~\eqref{Gammagamma}, \eqref{GammaWW}, \eqref{GammaZZ} and \eqref{Gammahh}. 

As shown in \cite{Hisano:2004ds}, the Sommerfeld effect is encoded, in the basis of the inert pair states $(H^0,H^0)$, $(A^0,A^0)$ and $(H^-,H^+)$, in a $3\times 3$ matrix $g(r)$, which satisfies the following second-order differential equation
\begin{eqnarray}
g''(r) + M_{H^0} \left(\dfrac{1}{4} M_{H^0} v^2 {1\!\!1} - V(r) \right) g(r) =0\;,
\label{SoDE}
\end{eqnarray}
where $v$ is the relative velocity of the initial state particles and $V(r)$ is given in Eq.~\eqref{potential}. The boundary conditions to solve the differential equation can be determined by analyzing the behavior of the solution $g(r)$ at $r= 0$ ant $r\rightarrow\infty$. At the origin,
\begin{eqnarray}
g(0)&=&{1\!\!1}\;.
\end{eqnarray}
On the other hand, for large values of $r$, the matrix $g(r)$ depends on the mass splitting between the pair states. If the mass splitting $\delta m_{ij}$ associated to the inert pairs $i$ and $j$ is smaller than the initial kinetic energy $\dfrac{1}{4}M_{H^0} v^2$, then there is enough energy to produce on-shell states of the corresponding pair and, therefore, the matrix element $g_{ij}(r)$ at infinity behaves as an out-going wave with momentum given, according to Eq.~(\ref{SoDE}), by 
\begin{eqnarray}
p_i= \sqrt{M_{H^0} \left(\dfrac{1}{4} M_{H^0} v^2 - V_{ii}(\infty) \right)} = \sqrt{M_{H^0} \left(\dfrac{1}{4} M_{H^0} v^2 - 2\,  \delta m_{ii}\right)}\;.
\end{eqnarray}
The corresponding boundary condition is
\begin{eqnarray}
\frac{d g_{ij}(r)}{dr} = i \, p_{i} \, g_{ij} (r)\,~~~\text{when~}r\rightarrow\infty\;.
\end{eqnarray}
In the opposite case, namely when $\delta m_{ii}> \dfrac{1}{4}M_{H^0} v^2$, there is not enough energy to produce on-shell states of the corresponding pair, and therefore the matrix elements $g_{ij}(r)$ decay exponentially at infinity. Hence
\begin{eqnarray}
g_{ij}(r) = 0\,,~~~\text{when~}r\rightarrow\infty\;.
\end{eqnarray}
The boundary conditions at $r=0$ and $r\rightarrow\infty$ then allow to find $g(r)$ by solving Eq.~\eqref{SoDE}. 

Once the solution is obtained,  the oscillating phases of $g(r)$ at large values of $r$ can be factorized by casting the matrix as
\begin{equation}
g(r) \to e^{ ir\sqrt{M_{H^0} \left(\frac{1}{4} M_{H^0} v^2 {1\!\!1} - 2\,  \delta m\right)}} d\;.
\end{equation}
where $d$ is a $3\times 3$ matrix which contains the non-perturbative enhancement factors due to the exchange of scalar and gauge bosons in the initial state.\footnote{An alternative method to calculate the Sommerfeld enhancement factors was proposed in Ref.~\cite{Garcia-Cely:2015dda} and consists in solving instead a differential equation for the matrix $h(r)=g'(r)g(r)^{-1}$ with appropriate boundary conditions, and which cures the numerical instabilities that plague the numerical solution of Eq.~\ref{SoDE}.}

Finally, the s-wave cross section for the annihilation of the pair $i$ into a final state $f$ can be determined from
\begin{equation}
\sigma v \left(i \to f\right) \Big|_{s-wave} = \frac{1}{N_{i}^2} (d\,\Gamma_f \, d^\dagger)_{ii}\,,
\label{SEsigmav}
\end{equation}
where $N_i=1/\sqrt{2}$ for initial states with identical particles, and $N_i=1$ otherwise.  
Notice that when the potential in Eq.~(\ref{SoDE}) is negligible then $d={1\!\!1}$, and therefore Eqs.~(\ref{SEsigmav2}) and (\ref{SEsigmav}) reduce to the standard expressions for calculating the cross section in the s-wave limit.

\section{Impact of the Sommerfeld Enhancement in the Early Universe}
\label{sec:AppendixC}
  
The DM thermal freeze-out occurs at a temperature $T_{FO}\sim M_{H^0}/20$. For masses greater than a few TeV, this corresponds to the era before the electroweak symmetry breaking, when the isospin is a good quantum number and the co-annihilating species are degenerate in mass. Due to this, the potential and annihilation matrices take a particular simple form, thus allowing to estimate the impact of the Sommerfeld enhancement in the early Universe~\cite{Cirelli:2009uv}. 

Let us first consider pairs of the co-annihilating species ${H^0, A^0, H^+, H^-}$. The subspace generated by such pairs can be decomposed into one self-conjugate isospin singlet $|m_I=0,I=0\rangle_{Y=0}$, one self-conjugate triplet $|m_I,I=1\rangle_{Y=0}$ and one isospin triplet $|m_I,I=1\rangle_{Y=1}$ and its corresponding complex conjugate $|m_I,I=1\rangle_{Y=-1}$. In terms of these states, the co-annihilating pairs with charge $Q=0$, $Q=1$ and $Q=2$ read
\begin{eqnarray}
Q&=&2 : \hspace{20pt}
 H^+\,H^+= |m_I=1, I=1\rangle_{Y=1}  
\label{eq:rotQ2} \\
Q&=&1 : \hspace{20pt}
\left(
\begin{array}{l}
 H^+\,H^0 \\
 H^+\,A^0\\
\end{array}
\right)
=
\left(
\begin{array}{ccccccc}
   \frac{1}{\sqrt{2}} & \frac{1}{\sqrt{2}} \\
  -\frac{i}{\sqrt{2}} & \frac{i}{\sqrt{2}} \\
\end{array}
\right)
\left(
\begin{array}{l}
 |m_I=0, I=1\rangle_{Y=1}\\
 |m_I=1, I=1\rangle_{Y=0}\\
\end{array}
\right)
\label{eq:rotQ1}\\
Q&=&0 : \hspace{20pt}
\left(
\begin{array}{l}
 H^0\,H^0\\
 A^0\,A^0\\
 H^+\,H^- \\
 H^0\, A^0 \\
\end{array}
\right)
=
\left(
\begin{array}{cccc}
  \frac{1}{2} & \frac{1}{2} & \frac{1}{2} & \frac{1}{2} \\
  \frac{1}{2} & \frac{1}{2} & -\frac{1}{2} & -\frac{1}{2} \\
  \frac{1}{\sqrt{2}} & -\frac{1}{\sqrt{2}} & 0 & 0 \\
  0&0&-\frac{i}{\sqrt{2}} & \frac{i}{\sqrt{2}} \\
\end{array}
\right)
\left(
\begin{array}{l}
 |m_I=0, I=0\rangle_{Y=0}\\
 |m_I=0, I=1\rangle_{Y=0}\\
 |m_I=-1, I=1\rangle_{Y=1}\\
 |m_I=1, I=1\rangle_{Y=-1}\\
\end{array}
\right)
\label{eq:rotQ0}
\end{eqnarray}
Because the isospin is a good quantum number, the non-relativistic potential is proportional to the identity in subspaces with definite isospin and hypercharge. Concretely,
\begin{eqnarray}
V \,|m_I, I\rangle_{Y} = V_{IY} |m_I, I\rangle_{Y}
\end{eqnarray}
This equation and the previous transformation  matrices allow to express the potential in the basis of co-annihilating pairs. Comparing the resulting potential for $Q=0$ with Eq.~\eqref{potential}, it is possible to solve for the potential $V_{IY}$, after neglecting the vev triggering the electroweak symmetry breaking (and therefore the mass splittings, the gauge boson masses and the scalar potential, which are all proportional to  $v_h$). The result reads:
\begin{equation}
V_{IY}=\frac{\alpha_{IY}}{r}\,,
\end{equation}
with
\begin{equation}
 \alpha_{I=0,Y=0} =- \frac{\left(1+2\,c_W^2\right) \, g^2}{16 \pi c_W^2 },\hspace{10pt}
 \alpha_{I=1,Y=1} = \frac{g^2}{16 \pi c_W^2 }, \hspace{10pt}
\alpha_{I=1,Y=0} =  \frac{\left(-1+2\,c_W^2\right)\,g^2}{16 \pi c_W^2 }
\end{equation}
An analagous procedure can be applied to the annihilation matrices. That is 
\begin{eqnarray}
\Gamma \,|m_I, I\rangle_{Y} = \Gamma_{IY} |m_I, I\rangle_{Y}
\end{eqnarray}
Once again by employing this and Eqs.~\eqref{eq:rotQ2}, \eqref{eq:rotQ1} and \eqref{eq:rotQ0}, we write the annihilation matrices in the basis of co-annihilating pairs. Furthermore by comparing the result with the addition of Eqs.~\eqref{Gammagamma}, \eqref{GammaWW}, \eqref{GammaZZ} and \eqref{Gammahh}, we solve for the annihilation matrix 
\begin{eqnarray}
 \Gamma_{I=0,Y=0} &= & \frac{1}{32\pi M^2_{H^0}} \left(\left( \frac{1-2\,c_W^2+4\,c_W^4}{2 \, c_W^4}\right)  g^4 + \left(2\lambda_3+\lambda_4\right)^2\right)\\
 \Gamma_{I=1,Y=1} &= & \frac{\lambda_5^2}{32\pi M^2_{H^0}}\\
 \Gamma_{I=1,Y=0} &= & \frac{1}{32\pi M^2_{H^0}} \left(\left( -1 +\frac{1}{c_W^2}\right)  g^4 + \lambda_4^2\right)  
\end{eqnarray}

In each subspace with definite hypercharge and isospin, the potential and the annihilation matrices take, neglecting the gauge boson masses, the simple form of a Coulomb potential, thus allowing to analytically  estimate the Sommerfeld enhancement in each subspace. Concretely, the total annihilation cross sections reads
\begin{eqnarray}
(\sigma v )_{eff} &=& \frac{2}{4^2} \sum_{I, Y}  (2I+1) \left(\frac{  \pi\alpha_{IY}/v }{e^{\pi\alpha_{IY}/v }-1}\right) \Gamma_{IY}\;,
\label{sigmaveff}
\end{eqnarray}
where the function $z/(e^z-1)$, with $z=\pi \alpha_{IY}/v$, is the well known  Sommerfeld enhancement factor associated to a Coulomb potential \cite{Cirelli:2007xd, Sommerfeld}. Besides, the factor of 2 in the numerator comes from the fact that the DM in the IDM is its own antiparticle, the symmetry factor $4^2$ is the total number of pairs that can be constructed from the co-annihilating species, and the factor $(2I+1)$ is the multiplicity of each isospin state

Lastly, taking the thermal average of this expression and  using the instantaneous freeze-out approximation, one can estimate the DM relic density as~\cite{Hambye:2009pw}
\begin{equation}
\Omega h^2 =  \frac{1.07\times 10^9\, \text{GeV}^{-1}}{g_\star^{1/2} M_{Pl} } \left(\int^\infty_{x_f} \frac{\langle \sigma v \rangle_{eff}}{x^2} dx \right)^{-1}\
\label{relicSymSU2},
\end{equation}
where the inverse freeze-out temperature $x_f=M_{H^0}/T_f$ can be found by solving
\begin{equation}
x_f = \log \left( \frac{0.038 \,M_{Pl} M_{H^0}  x_f^{1/2} \langle \sigma v \rangle_{eff} }{g_\star^{1/2}}\right)\,.
\end{equation}

Remarkably, Eq.~\eqref{relicSymSU2} can be used to calculate the relic density even where there is no Sommerfeld enhancement, by taking $\alpha_{IY}\to 0$. In this limit, Eq.~\eqref{sigmaveff} reduces to
\begin{eqnarray}
(\sigma v )_{eff} &\simeq& \frac{2.4+ 5.2 \left[ (2\lambda_3 + \lambda_4)^2 + 3\lambda_4^2 + 6\lambda_5^2\right]}{\left(M_{H^0}/530 \text{ GeV} \right)^2}\times 10^{-26} 
\text{cm}^3/\text{s}. 
\label{sigmaveff22}
\end{eqnarray}

From this, we find a deviation of at most 10\% from the result of the scan of Section \ref{sec:DM-annihilation}, which was derived using  micrOMEGAs 3.1~\cite{Belanger:2013oya}. We also find that the difference becomes stronger when the mass splitting is very large, as expected from the fact that in this regime the approximation of taking the $SU(2)_L$ symmetric limit is worst.  When the Sommerfeld enhancement is taken into account, the disagreement with respect to perturbative result is at most of 30\%, in remarkable agreement with what has been found for Higgsino DM~\cite{Beneke:2014hja, Hisano:2006nn}, another $SU(2)_L$ doublet candidate. We thus conclude the the Sommerfeld effect in the early Universe is less dramatic than in the galactic center.

For a given set of quartic couplings, using Eq.~\eqref{sigmaveff22} and  $\Omega_{DM} \simeq \frac{3\times 10^{-27} \text{cm}^3\text{s}^{-1}}{\langle \sigma v\rangle_{eff}} \simeq 0.12$,  it is straightforward to approximately find  the mass $M_{H^0}$ corresponding to the observed relic abundance. In fact, with the constraints of Appendix~\ref{sec:AppendixA}, we find that the maximal allowed $M_{H^0}$ is about $22.4$~TeV (e.g.\ at the point $\lambda_3 = 4\pi$, $\lambda_4=\lambda_5 \simeq- 2.4\pi$ and $\lambda_2 =4\pi$).

\bibliographystyle{JHEP}
\bibliography{text}

\end{document}